\documentclass[pdflatex, referee, sn-mathphys]{sn-jnl}

\usepackage{comment}
\usepackage{bm}
\usepackage{mathtools}
\usepackage{fancyhdr}

\theoremstyle{thmstyleone}%
\theoremstyle{thmstyletwo}%

\theoremstyle{thmstylethree}%

\begin{document}

\title[ ]{Correspondence between open bosonic systems and stochastic differential equations}


\author*[1]{\fnm{Alexander} \sur{Engel}}\email{alen3220@colorado.edu}

\author[1,2]{\fnm{Scott E.} \sur{Parker}} 

\affil[1]{\orgdiv{Department of Physics}, \orgname{University of Colorado}, \orgaddress{\city{Boulder}, \state{Colorado} \postcode{80309}, \country{USA}}} 

\affil[2]{\orgdiv{Renewable and Sustainable Energy Institute}, \orgname{University of Colorado}, \orgaddress{\city{Boulder}, \state{Colorado} \postcode{80309}, \country{USA}}} 

\abstract{Bosonic mean-field theories can approximate the dynamics of systems of $n$ bosons provided that $n \gg 1$. We show that there can also be an exact correspondence at finite $n$ when the bosonic system is generalized to include interactions with the environment and the mean-field theory is replaced by a stochastic differential equation. When the $n \to \infty$ limit is taken, the stochastic terms in this differential equation vanish, and a mean-field theory is recovered. Besides providing insight into the differences between the behavior of finite quantum systems and their classical limits given by $n \to \infty$, the developed mathematics can provide a basis for quantum algorithms that solve some stochastic nonlinear differential equations. We discuss conditions on the efficiency of these quantum algorithms, with a focus on the possibility for the complexity to be polynomial in the log of the stochastic system size. A particular system with the form of a stochastic discrete nonlinear Schr\"{o}dinger equation is analyzed in more detail.}



\maketitle

\fancyhf{}
\renewcommand{\headrulewidth}{0pt}
\lfoot{\bf \footnotesize This preprint has not undergone peer review or any post-submission improvements or corrections. The Version of Record of this article is published in The European Physical Journal Plus, and is available online at \url{https://doi.org/10.1140/epjp/s13360-023-04205-9}.}
\thispagestyle{fancy}

\section{Introduction}

The dynamics of small quantum systems, i.e., systems comprising only a few quantum objects, can be predicted through direct simulation on classical computers. Additionally, the dynamics of very large quantum systems are often described by approximations, such as mean-field theories, that consider the limit of infinite system size. Still, this leaves a wide range of intermediate system sizes that are difficult to study. Quantum computers lie within this range, and by using them to run Hamiltonian simulation algorithms (e.g., \cite{Berry2007, Low2017, Qubitization}), the dynamics of other intermediate-size quantum systems can be efficiently simulated.

Other quantum algorithms allow quantum computers to solve problems unrelated to quantum physics with speedups relative to existing classical algorithms. These include quantum linear systems algorithms \cite{Harrow2009, QLSA2017}, quantum algorithms for linear differential equations \cite{ODE2017, Childs2021}, and quantum algorithms for linear simulations of classical waves, fluids, and plasmas \cite{Engel2019, Costa2019, Todorova, Dodin2021, MechanicalReview2022}. Applying quantum computers to solve nonlinear problems is less straightforward, but a variety of approaches have been proposed and studied \cite{PoP, Dissipative, HomotopyPerturbationMethod, Variational, Variational2021, MechanicalReview2022, Joseph2020, Dodin2021, NonlinearPDE2022, QuantumFluid}.

Every quantum algorithm can be viewed as a prescription for the evolution of the quantum system that is the quantum computer such that the mathematics of this linear, unitary evolution generates a solution to the target problem. Thus, for nonlinear simulation problems, a natural strategy is to look for mappings that relate nonlinear dynamics to the unitary evolution of many-body quantum systems. Mean-field theory provides these types of mappings, and using mean-field approximations as the basis of a quantum algorithm for simulating general nonlinear dynamics was proposed in \cite{Lloyd2020}.

However, when the system size is not extremely large, the accuracy of mean-field approximations is suspect. For instance, by applying a discretized Gross--Pitaevskii equation to the problem of state discrimination, Childs and Young \cite{BEC2016} deduced that this mean-field approximation must break down in time $t = \mathcal{O}(g^{-1}\log n)$, where $n$ is the number of particles, and $g$ quantifies the strength of the nonlinearity in the Gross--Pitaevskii equation. For comparison, error bounds obtained for bosonic mean-field theories \cite{Rodnianski2009, Chen2011, Paul2019} imply accuracy out to $t = \Omega(\log n)$. So it appears that using a number of particles exponential in the total simulation time is necessary and sufficient for mean-field approximation accuracy. But then, even macroscopic quantum systems should be too small to apply mean-field approximations for moderate simulation times.

We study mappings that are valid for arbitrary $n$, thus allowing for a better understanding of the behavior of some intermediate-size quantum systems. This is relevant both for assessing the accuracy of mean-field approximations for real, intermediate-size quantum systems and for bounding the errors of any quantum algorithms that are based on mean-field approximations. We begin by introducing a representation of density matrices in Sect.~\ref{sec:hyp} that allows for mapping the evolution of open bosonic systems to partial differential equations. Next, in Sect.~\ref{sec:consde}, we connect those partial differential equations to stochastic differential equations. In Sect.~\ref{eq:stqa} we investigate the efficiency of an approach for simulating the open bosonic systems on quantum computers, which provides a potential basis for quantum algorithms for stochastic differential equations; and we introduce a stochastic discrete nonlinear Schr\"{o}dinger equation as an example. Concluding remarks are then given in Sect.~\ref{sec:conc}.

\section{Density matrix evolution as a partial differential equation}\label{sec:hyp}

\subsection{Hyperspherical representation}
Let
\begin{equation}\label{eq:hyp}
\rho(t) = \int_{\vert{\bf z}\vert=1} f({\bf z},t) \left(\vert {\bf z} \rangle \langle {\bf z} \vert\right)^{\otimes n} d \text{Re}({\bf z}) d\text{Im}({\bf z})
\end{equation}
define a mapping from a real-valued distribution $f({\bf z}, t)$ to a complex operator $\rho(t)$. The notation $\vert {\bf z} \rangle$ for a complex, $N$-component vector ${\bf z}$ is taken to mean a quantum state
\begin{equation}\label{eq:zstate}
\vert {\bf z} \rangle \equiv \sum_{j=0}^{N-1} z_j \vert j \rangle,
\end{equation}
which is considered to be a single-particle state. Explicit index limits will be omitted when they can be inferred; e.g., (\ref{eq:zstate}) is $\vert {\bf z} \rangle \equiv \sum_j z_j \vert j \rangle$. Also, $d \text{Re}({\bf z})$ is shorthand for $\prod_j d \text{Re}(z_j)$ while $d\text{Im}({\bf z})$ means $\prod_j d \text{Im}(z_j)$. By construction, $\rho(t)$ is bosonic and Hermitian since the integrand in (\ref{eq:hyp}) is a bosonic, Hermitian operator over $n$ particles. We call an operator bosonic when it is non-zero only within the bosonic subspace, which is a property of density matrices of bosonic systems. Now, suppose that $f({\bf z}, t)$ satisfies
\begin{equation}\label{eq:unit}
\int_{\vert{\bf z}\vert=1} f({\bf z},t) d \text{Re}({\bf z}) d\text{Im}({\bf z}) = 1.
\end{equation}
Then
\begin{equation}
\begin{split}
\text{Tr} [\rho(t)] &= \int_{\vert{\bf z}\vert=1} f({\bf z},t) \text{Tr}\left[ \left(\vert {\bf z} \rangle \langle {\bf z} \vert\right)^{\otimes n}\right] d \text{Re}({\bf z}) d\text{Im}({\bf z}) \\
 &= \int_{\vert{\bf z}\vert=1} f({\bf z},t) d \text{Re}({\bf z}) d\text{Im}({\bf z}) \\
 &= 1.
\end{split}
\end{equation}

For $\rho(t)$ to be a bosonic density matrix, it must also be positive semi-definite. In our applications, this will be ensured through the initialization and evolution of $\rho(t)$. In particular, for an initially $\delta$-distributed $f({\bf z},t)$,
\begin{equation}\label{eq:rho0}
\rho(0) = \left(\vert {\bf z}_0 \rangle \langle {\bf z}_0 \vert\right)^{\otimes n}
\end{equation}
for some normalized ${\bf z}_0$, which is clearly positive semi-definite. Physical density matrix evolution, such as that due to the von Neumann equation or, more generally, a master equation in Lindblad form, is guaranteed to preserve the positive semi-definiteness of the density matrix. So, if $f({\bf z},t)$ is evolved such that $\rho(t)$ satisfies a physical evolution equation, then $\rho(t)$ is a bosonic density matrix for all $t$. The problem of finding suitable $f({\bf z},t)$ evolution is covered in Sect.~\ref{sec:mapev}.

The initial condition of (\ref{eq:rho0}) can be significantly generalized: any $\rho(0)$ obtained [through (\ref{eq:hyp})] from $f({\bf z},0)$ with $f({\bf z},0) \geq 0$ is positive semi-definite. This follows from convexity of the set of positive semi-definite matrices. Additionally, since these $\rho(0)$ are linear combinations of product density matrices with positive weights, they are separable. In other words, the bosonic particles are not entangled with one another. Of course, generic unitary evolution can create entanglement. Therefore, $f({\bf z},t)$ with $f({\bf z},t) \geq 0$ cannot represent generic unitary evolution. But the $f({\bf z},t)$ in (\ref{eq:hyp}) is allowed to have both signs, which makes this representation more general.

Further analysis is facilitated by the introduction of Wirtinger derivatives:
\begin{equation}
\bm{\nabla} \equiv \frac12 \left( \frac{\partial}{\partial {\bf x}} - i \frac{\partial}{\partial {\bf y}}\right),
\end{equation}
where ${\bf x} = \text{Re}({\bf z})$ and ${\bf y} = \text{Im}({\bf z})$. Wirtinger derivatives satisfy the commutation relations
\begin{align}\label{eq:Wirtrel}
\left[\nabla_j, z_k\right] &= \delta_{jk}, & \left[\nabla^*_j, z_k\right] &= 0,
\end{align}
and they can be applied to pick out specific operator components. For instance,
\begin{equation}
\nabla_j \nabla^*_k \vert {\bf z} \rangle \langle {\bf z} \vert = \vert j \rangle \langle k \vert.
\end{equation}
More generally, for vectors ${\bf n}^l$ and ${\bf n}^r$ of non-negative integers that each sum to $n$,
\begin{equation}\label{eq:ops}
\left(\prod_{j=0}^{N-1} \nabla^{n^l_j}_j\right) \left(\prod_{j=0}^{N-1} \left(\nabla^*\right)^{n^r_j}_j \right) \left(\vert {\bf z} \rangle \langle {\bf z} \vert\right)^{\otimes n} \propto \vert {\bf n}^l \rangle \langle {\bf n}^r \vert,
\end{equation}
where $\vert {\bf n}^l \rangle$ and $\vert {\bf n}^r \rangle$ are bosonic states with definite numbers of particles given by the components of ${\bf n}^l$ and ${\bf n}^r$, respectively. Such states comprise a basis for all $n$-particle bosonic states, and so, the operators in (\ref{eq:ops}) comprise a basis for all $n$-particle bosonic operators. Hermitian operators in particular can be constructed using only the Hermitian components,
\begin{multline}\label{eq:hops}
\frac12 \left[\left(\prod_{j=0}^{N-1} \nabla^{n^l_j}_j\right) \left(\prod_{j=0}^{N-1} \left(\nabla^*\right)^{n^r_j}_j \right) \left(\vert {\bf z} \rangle \langle {\bf z} \vert\right)^{\otimes n} + \text{h.c.} \right] \\
\begin{split}&= \text{Re}\left[\left(\prod_{j=0}^{N-1} \nabla^{n^l_j}_j\right) \left(\prod_{j=0}^{N-1} \left(\nabla^*\right)^{n^r_j}_j \right)\right] \left(\vert {\bf z} \rangle \langle {\bf z} \vert\right)^{\otimes n} \\
&\propto \vert {\bf n}^l \rangle \langle {\bf n}^r \vert + \vert {\bf n}^r \rangle \langle {\bf n}^l \vert,\end{split}
\end{multline}
where $\text{h.c.}$ is shorthand for the Hermitian conjugate of the operator to its left. Note that this Hermitian conjugation is over the space of bosonic operators, and so it treats the Wirtinger derivatives as scalars. (\ref{eq:hops}) implies that any bosonic, Hermitian, $n$-particle operator $\rho$ is representable as
\begin{equation}\label{eq:genrho}
\rho = Q \left(\vert {\bf z} \rangle \langle {\bf z} \vert\right)^{\otimes n},
\end{equation}
where $Q$ is a real, linear combination of order-$2n$ derivatives with respect to the variables in ${\bf x}$ and ${\bf y}$. These derivatives eliminate all factors of ${\bf z}$ in (\ref{eq:genrho}) such that $\rho$ does not depend on ${\bf z}$.

Derivatives can be sourced from distributions. For instance, given a distribution $h(x)$ that satisfies
\begin{align}
\int_{-\infty}^\infty h(x) dx &= 1, & \lim_{x \to \pm \infty} x h(x) &= 0,
\end{align}
$h'(x) = dh(x)/dx$ can be used to apply a derivative to a quantum state in an integrand:
\begin{equation}
-\int_{-\infty}^\infty h'(x_j) \vert {\bf z} \rangle dx_j = \int_{-\infty}^\infty h(x_j) \left[ \frac{d}{dx_j} \vert {\bf z} \rangle \right] dx_j = \int_{-\infty}^\infty h(x_j) \vert j \rangle dx_j = \vert j \rangle.
\end{equation}
However, this relies on integration by parts, which is less straightforward over the integration domain of (\ref{eq:hyp}). Consequently, it is useful to note that (\ref{eq:hyp}) is no less general than
\begin{equation}\label{eq:uncstr}
\rho(t) = \int g({\bf z},t) \left(\vert {\bf z} \rangle \langle {\bf z} \vert\right)^{\otimes n} d \text{Re}({\bf z}) d\text{Im}({\bf z}),
\end{equation}
where the integration domain is now the full space, and $g({\bf z},t)$ is any distribution that vanishes sufficiently fast as $\vert {\bf z} \vert \to \infty$ to ensure that (\ref{eq:uncstr}) is well defined. Each point with $\vert {\bf z}\vert \neq 1$ can be replaced with a point on the hypersphere based on
\begin{equation}\label{eq:pointmap}
\left(\vert {\bf z} \rangle \langle {\bf z} \vert\right)^{\otimes n} = \vert {\bf z} \vert^{2n} \left(\left\vert \frac{\bf z}{\vert{\bf z}\vert} \right\rangle \left\langle \frac{\bf z}{\vert{\bf z}\vert} \right\vert\right)^{\otimes n}.
\end{equation}
Therefore, the integrand in (\ref{eq:uncstr}) can be projected onto the hypersphere (i.e., onto $\vert {\bf z} \vert = 1$) to obtain an expression in the form of (\ref{eq:hyp}) with $f({\bf z},t)$ being real valued provided that $g({\bf z},t)$ is real valued.

Now, (\ref{eq:genrho}) can be expressed in the form of (\ref{eq:hyp}) as follows. First,
\begin{equation}\label{eq:cvs1}
\begin{split}
\rho &= \int \rho \, h({\bf z}) d \text{Re}({\bf z}) d\text{Im}({\bf z}),\\
&= \int \left[Q \left(\vert {\bf z} \rangle \langle {\bf z} \vert\right)^{\otimes n}\right] h({\bf z}) d \text{Re}({\bf z}) d\text{Im}({\bf z}),
\end{split}
\end{equation}
where $h({\bf z})$ is any real-valued distribution that integrates to one and becomes negligible as $\vert {\bf z} \vert \to \infty$. Next, by applying integration by parts $2n$ times, all derivatives in $Q$ can be transferred over to obtain
\begin{equation}
\rho = \int \left[Q h({\bf z}) \right] \left(\vert {\bf z} \rangle \langle {\bf z} \vert\right)^{\otimes n} d \text{Re}({\bf z}) d\text{Im}({\bf z}).
\end{equation}
A $t$ dependence can also be introduced, showing up in $\rho(t)$ and $Q(t)$. Then $\rho(t)$ is expressed in the form of (\ref{eq:uncstr}) with
\begin{equation}\label{eq:gzw}
g({\bf z}, t) = Q(t) h({\bf z}),
\end{equation}
and projection onto $\vert {\bf z} \vert = 1$ converts this into the form of (\ref{eq:hyp}).

So (\ref{eq:hyp}) can represent general Hermitian operators on $n$ bosons. Of course, if (\ref{eq:unit}) is assumed, only those with unit trace are possible, which is appropriate for a density matrix representation. Either way, this representation is overcomplete. For example, different choices for the $h({\bf z})$ distribution in (\ref{eq:cvs1}) can produce different $f({\bf z}, t)$ that represent the same $\rho(t)$. Due to this overcompleteness, the $f({\bf z}, t)$ evolution is not uniquely determined by $\rho(t)$. The $f({\bf z}, t)$ evolution that shall now be derived is just a particularly straightforward choice.

\subsection{Mapping of evolution terms}\label{sec:mapev}
The evolution of $\rho$ can be expressed generically as
\begin{equation}\label{eq:rhosup}
\dot{\rho} = \mathcal{L}[\rho],
\end{equation}
where $\mathcal{L}[\cdot]$ is a linear superoperator, and the time dependence of $\rho$ is implicit. Taking the time derivative of (\ref{eq:hyp}) and applying (\ref{eq:rhosup}) gives
\begin{equation}\label{eq:eveq}
\int_{\vert{\bf z}\vert=1} \frac{\partial f({\bf z},t)}{\partial t} \left(\vert {\bf z} \rangle \langle {\bf z} \vert\right)^{\otimes n} d \text{Re}({\bf z}) d\text{Im}({\bf z}) = \int_{\vert{\bf z}\vert=1} f({\bf z},t) \mathcal{L}\left[\left(\vert {\bf z} \rangle \langle {\bf z} \vert\right)^{\otimes n}\right] d \text{Re}({\bf z}) d\text{Im}({\bf z}).
\end{equation}
Next, an operator $L$ is constructed out of ${\bf z}$, ${\bf z}^*$, $\bm{\nabla}$, and $\bm{\nabla}^*$ such that
\begin{equation}\label{eq:replace}
\mathcal{L}\left[\left(\vert {\bf z} \rangle \langle {\bf z} \vert\right)^{\otimes n}\right] = L\left(\vert {\bf z} \rangle \langle {\bf z} \vert\right)^{\otimes n}
\end{equation}
for all ${\bf z}$. For this step we restrict to evolution that is expressible with polynomials of bosonic creation and annihilation operators, written as $\hat{\bf a}^\dagger$ and $\hat{\bf a}$ respectively. It is also helpful to introduce coherent states,
\begin{equation}\label{eq:coh}
\vert \psi_{\bf z} \rangle \equiv \exp\left({\bf z} \cdot \hat{\bf a}^\dagger\right) \vert 0 \rangle,
\end{equation}
where $\vert 0 \rangle$ is the state with no particles. The properties of coherent states allow for this simple derivation of a useful relation:
\begin{equation}\label{eq:relcoh}
\begin{split}
\left(\prod_i \hat{a}^\dagger_{\beta_i}\right) \left(\prod_i \hat{a}_{\alpha_i}\right)\vert \psi_{\bf z} \rangle &= 
\left(\prod_i \hat{a}^\dagger_{\beta_i}\right) \left(\prod_i z_{\alpha_i}\right) \vert \psi_{\bf z} \rangle \\
&= \left(\prod_i z_{\alpha_i}\right) \left(\prod_i \hat{a}^\dagger_{\beta_i}\right) \vert \psi_{\bf z} \rangle \\
&= \left(\prod_i z_{\alpha_i}\right) \left(\prod_i \nabla_{\beta_i}\right) \vert \psi_{\bf z} \rangle
\end{split}
\end{equation}
for multi-indices $\alpha$ and $\beta$. Suppose from now on that $\alpha$ and $\beta$ have the same length. Then the operators on both sides of
\begin{equation}\label{eq:compz}
\left(\prod_i \hat{a}^\dagger_{\beta_i}\right) \left(\prod_i \hat{a}_{\alpha_i}\right)\vert \psi_{\bf z} \rangle = \left(\prod_i z_{\alpha_i}\right) \left(\prod_i \nabla_{\beta_i}\right) \vert \psi_{\bf z} \rangle
\end{equation}
are particle conserving; the ones on the right never alter the number of particles regardless. Further, since the $n$-particle component of (\ref{eq:coh}) is $\propto \vert {\bf z} \rangle^{\otimes n}$, (\ref{eq:compz}) implies that
\begin{equation}\label{eq:map0}
\left(\prod_i \hat{a}^\dagger_{\beta_i}\right) \left(\prod_i \hat{a}_{\alpha_i}\right)\vert {\bf z} \rangle^{\otimes n} = \left(\prod_i z_{\alpha_i}\right) \left(\prod_i \nabla_{\beta_i}\right)\vert {\bf z} \rangle^{\otimes n},
\end{equation}
which provides a starting point for determining the mapping from $\mathcal{L}[\cdot]$ to $L$. Operators to the left of $(\vert {\bf z} \rangle \langle {\bf z} \vert)^{\otimes n}$ in $\mathcal{L}[(\vert {\bf z} \rangle \langle {\bf z} \vert)^{\otimes n}]$ can be replaced using (\ref{eq:map0}), but the raising operators have to be to the left of the lowering operators. However, it is straightforward to extend (\ref{eq:map0}) to operators in any order. Note that $\hat{a}_{\alpha_i}$ is converted to $z_{\alpha_i}$, while $\hat{a}^\dagger_{\beta_i}$ is converted to $\nabla_{\beta_i}$. This is associated with a sign change of the commutation relations, since
\begin{equation}\label{eq:sgdf}
\begin{split}
[\hat{a}_j, \hat{a}^\dagger_k] &= \delta_{jk},\\
[z_j, \nabla_k] &= -\delta_{jk}.
\end{split}
\end{equation}
This sign change is consistent provided that the mapping includes an order reversal:
\begin{multline}\label{eq:mapleft}
\left(\prod_i \hat{a}^\dagger_{\alpha^0_i}\right) \left(\prod_i \hat{a}_{\alpha^1_i}\right) ... \left(\prod_i \hat{a}^\dagger_{\alpha^{m-1}_i}\right) \left(\prod_i \hat{a}_{\alpha^m_i}\right) \vert {\bf z} \rangle^{\otimes n} \\
= \left(\prod_i z_{\alpha^m_i}\right) \left(\prod_i \nabla_{\alpha^{m-1}_i}\right) ... \left(\prod_i z_{\alpha^1_i}\right) \left(\prod_i \nabla_{\alpha^0_i}\right) \vert {\bf z} \rangle^{\otimes n},
\end{multline}
where it is assumed that the operator acting on $\vert {\bf z} \rangle^{\otimes n}$ on the left side of (\ref{eq:mapleft}) conserves the total number of particles, which constrains the $\alpha^k$ multi-indices. In words, this mapping consists of replacing $\hat{a}_j$ with $z_j$, replacing $\hat{a}^\dagger_j$ with $\nabla_j$, and reversing the order of all terms. Then, for example, when two adjacent terms are exchanged in both operator forms, the associated commutator terms agree because the sign difference in (\ref{eq:sgdf}) cancels with the sign difference due to the ordering being opposite. Swaps of adjacent terms can generate any permutation, so this proves general agreement.

It is still necessary to derive the mapping for operators that are to the right of $(\vert {\bf z} \rangle \langle {\bf z} \vert)^{\otimes n}$ in $\mathcal{L}[(\vert {\bf z} \rangle \langle {\bf z} \vert)^{\otimes n}]$. This is related to the previous result since
\begin{equation}
\langle {\bf z} \vert^{\otimes n} \hat{M} = \left(\hat{M}^\dagger \vert {\bf z} \rangle^{\otimes n}\right)^\dagger
\end{equation}
for any operator $\hat{M}$. The previous mapping just needs to be applied to $\hat{M}^\dagger$ followed by Hermitian conjugation, which changes the differential operator only through complex conjugation. The result is
\begin{multline}\label{eq:mapright}
\langle {\bf z} \vert^{\otimes n} \left(\prod_i \hat{a}^\dagger_{\alpha^0_i}\right) \left(\prod_i \hat{a}_{\alpha^1_i}\right) ... \left(\prod_i \hat{a}^\dagger_{\alpha^{m-1}_i}\right) \left(\prod_i \hat{a}_{\alpha^m_i}\right) \\
= \left(\prod_i z^*_{\alpha^0_i}\right) \left(\prod_i \nabla^*_{\alpha^1_i}\right) ... \left(\prod_i z^*_{\alpha^{m-1}_i}\right) \left(\prod_i \nabla^*_{\alpha^m_i}\right) \langle {\bf z} \vert^{\otimes n}.
\end{multline}
In words, the mapping for terms to the right is given by replacing $\hat{a}^\dagger_j$ with $z^*_j$ and replacing $\hat{a}_j$ with $\nabla^*_j$; the original operator order is maintained.

One important property of the differential operators is that the separation between operators that act on the left and operators that act on the right is preserved. Specifically, the Wirtinger derivatives satisfy $\nabla_j z^*_k = 0$, and so,
\begin{align}
\nabla_j \langle {\bf z} \vert^{\otimes n} &= 0, & \nabla^*_j \vert {\bf z} \rangle^{\otimes n} = 0.
\end{align}
Therefore, the differential operators from (\ref{eq:mapleft}) and (\ref{eq:mapright}) can both be applied to $(\vert {\bf z} \rangle \langle {\bf z} \vert)^{\otimes n}$ without generating extra terms. Additionally, since $\left[\nabla^*_j, z_k\right] = 0$, the differential operators in (\ref{eq:mapleft}) commute with the differential operators in (\ref{eq:mapright}). This makes it straightforward to combine the previous results to obtain
\begin{multline}\label{eq:mapfull}
\left[\left(\prod_i \hat{a}^\dagger_{\alpha^0_i}\right) \left(\prod_i \hat{a}_{\alpha^1_i}\right) ... \left(\prod_i \hat{a}^\dagger_{\alpha^{m-1}_i}\right) \left(\prod_i \hat{a}_{\alpha^m_i}\right)\right] \left(\vert {\bf z} \rangle \langle {\bf z} \vert\right)^{\otimes n} \\ \times \left[\left(\prod_i \hat{a}^\dagger_{\beta^0_i}\right) \left(\prod_i \hat{a}_{\beta^1_i}\right) ... \left(\prod_i \hat{a}^\dagger_{\beta^{m-1}_i}\right) \left(\prod_i \hat{a}_{\beta^m_i}\right)\right] \\
= \left[\left(\prod_i z_{\alpha^m_i}\right) \left(\prod_i \nabla_{\alpha^{m-1}_i}\right) ... \left(\prod_i z_{\alpha^1_i}\right) \left(\prod_i \nabla_{\alpha^0_i}\right)\right] \\
\times \left[\left(\prod_i z^*_{\beta^0_i}\right) \left(\prod_i \nabla^*_{\beta^1_i}\right) ... \left(\prod_i z^*_{\beta^{m-1}_i}\right) \left(\prod_i \nabla^*_{\beta^m_i}\right)\right] \left(\vert {\bf z} \rangle \langle {\bf z} \vert\right)^{\otimes n}.
\end{multline}
Given any $\mathcal{L}[\cdot]$ involving particle-conserving polynomials of bosonic creation and annihilation operators applied to one or both sides of its argument, the $L$ operator in (\ref{eq:replace}) is obtained by applying (\ref{eq:mapfull}) to each of the terms in $\mathcal{L}[\cdot]$, which works because this mapping between operator types is linear.

Here are a few important examples of this operator mapping. For non-interacting evolution, the bosonic Hamiltonian can be expressed as
\begin{equation}\label{eq:H1PI}
\hat{H} = H_{jk} \hat{a}_j^\dagger \hat{a}_k,
\end{equation}
where we employ the Einstein summation convention. Then, with $\hbar=1$,
\begin{equation}
\begin{split}
\mathcal{L}\left[\left(\vert {\bf z} \rangle \langle {\bf z} \vert\right)^{\otimes n}\right] &= -i H_{jk} \left[\hat{a}_j^\dagger \hat{a}_k \left(\vert {\bf z} \rangle \langle {\bf z} \vert\right)^{\otimes n} - \left(\vert {\bf z} \rangle \langle {\bf z} \vert\right)^{\otimes n} \hat{a}_j^\dagger \hat{a}_k \right] \\
&= -i H_{jk} \left( z_k \nabla_j - z^*_j \nabla^*_k \right) \left(\vert {\bf z} \rangle \langle {\bf z} \vert\right)^{\otimes n},
\end{split}
\end{equation}
and so,
\begin{equation}\label{eq:L1PI}
\begin{split}
L &= -i H_{jk} \left( z_k \nabla_j - z^*_j \nabla^*_k \right) \\
&= 2\text{Re}\left(-i H_{jk} z_k \nabla_j\right),
\end{split}
\end{equation}
where the second equality follows from $\hat{H} = \hat{H}^\dagger$ and shows that $L$ is real. Next, two-particle interactions can be expressed with a Hamiltonian
\begin{equation}\label{eq:H2PI}
\hat{H} = H_{jklm} \hat{a}_j^\dagger \hat{a}^\dagger_k \hat{a}_l \hat{a}_m,
\end{equation}
and the mapping gives
\begin{equation}\label{eq:L2PI}
\begin{split}
L &= -i H_{jklm} \left( z_m z_l \nabla_k \nabla_j - z^*_j z^*_k \nabla^*_l \nabla^*_m \right)\\
&= 2 \text{Re} \left( -i H_{jklm} z_m z_l \nabla_k \nabla_j \right),
\end{split}
\end{equation}
where again, the second equality follows from $\hat{H} = \hat{H}^\dagger$.

Evolution terms associated with open quantum systems can also be mapped. For instance, evolution of the form
\begin{equation}\label{eq:evX}
\dot{\rho} = - \left[ \hat{X}, \left[ \hat{X}, \rho \right]\right] = 2 \hat{X} \rho \hat{X} - \hat{X}^2 \rho - \rho \hat{X}^2
\end{equation}
for Hermitian $\hat{X}$ is allowed in Lindblad master equations. This evolution can be understood as the result of an unconditional continuous measurement process with a measurement operator proportional to $\hat{X}$ \cite{ContinuousMeasurements1987, ContinuousMeasurements2020}. This essentially means that the original system is interacting with the environment, and the experimenter does not keep track of the specific effects that this interaction causes. Now, suppose that $\hat{X}$ takes the simple form of
\begin{equation}\label{eq:formX}
\hat{X} = X_{jk} \hat{a}^\dagger_j \hat{a}_k.
\end{equation}
Then the evolution operator in (\ref{eq:evX}) maps to
\begin{equation}\label{eq:LX}
\begin{split}
L &= 2 X_{jk} z_k \nabla_j X_{lm} z^*_l \nabla^*_m - X_{jk} z_k \nabla_j X_{lm} z_m \nabla_l - X_{jk} z^*_j \nabla^*_k X_{lm} z^*_l \nabla^*_m \\
&= \left[2 \text{Im}\left( X_{jk} z_k \nabla_j \right) \right]^2,
\end{split}
\end{equation}
where the second equality follows from $\hat{X} = \hat{X}^\dagger$.

The next step in deriving the evolution of $f({\bf z}, t)$ is to insert (\ref{eq:replace}) into (\ref{eq:eveq}), yielding
\begin{equation}\label{eq:nextds}
\int_{\vert{\bf z}\vert=1} \frac{\partial f({\bf z},t)}{\partial t} \left(\vert {\bf z} \rangle \langle {\bf z} \vert\right)^{\otimes n} d \text{Re}({\bf z}) d\text{Im}({\bf z}) = \int_{\vert{\bf z}\vert=1} f({\bf z},t) \left[L \left(\vert {\bf z} \rangle \langle {\bf z} \vert\right)^{\otimes n}\right] d \text{Re}({\bf z}) d\text{Im}({\bf z}).
\end{equation}
Unfortunately, the $\vert{\bf z}\vert=1$ integration domain complicates the task of moving the derivatives in $L$ over to $f({\bf z},t)$. This difficulty can be sidestepped by switching to a form without the $\vert{\bf z}\vert=1$ constraint, similar to (\ref{eq:uncstr}). In particular, introduce $g({\bf z}, t)$ such that
\begin{equation}\label{eq:ucnmp}
\int \frac{g({\bf z},t)}{\vert{\bf z}\vert^{2n}} \left(\vert {\bf z} \rangle \langle {\bf z} \vert\right)^{\otimes n} d \text{Re}({\bf z}) d\text{Im}({\bf z}) = \int_{\vert{\bf z}\vert=1} f({\bf z},t) \left(\vert {\bf z} \rangle \langle {\bf z} \vert\right)^{\otimes n} d \text{Re}({\bf z}) d\text{Im}({\bf z}).
\end{equation}
For any $t$, a $g({\bf z},t)$ distribution satisfying (\ref{eq:ucnmp}) can easily be constructed from $f({\bf z},t)$:
\begin{equation}\label{eq:extension}
g({\bf z},t) = f({\bf z},t) \delta(\vert{\bf z}\vert - 1).
\end{equation}
Conversely, given some $g({\bf z},t)$, an $f({\bf z},t)$ distribution satisfying (\ref{eq:ucnmp}) can be constructed by projection onto $\vert{\bf z}\vert=1$. The purpose of the $\vert{\bf z}\vert^{-2n}$ factor included in (\ref{eq:ucnmp}) is to simplify this projection step. Specifically, this cancels the $\vert{\bf z}\vert^{2n}$ factor in (\ref{eq:pointmap}), which makes the projection step independent of $n$:
\begin{equation}\label{eq:projection}
f({\bf z},t) = \int_0^\infty g(r {\bf z}, t) dr.
\end{equation}
Now, when the steps leading to (\ref{eq:nextds}) are redone with the left side of (\ref{eq:ucnmp}) in place of the right side, the result is
\begin{equation}
\int \frac{\partial g({\bf z},t)}{\partial t} \frac{\left(\vert {\bf z} \rangle \langle {\bf z} \vert\right)^{\otimes n}}{\vert{\bf z}\vert^{2n}} d \text{Re}({\bf z}) d\text{Im}({\bf z}) = \int \frac{g({\bf z},t)}{\vert{\bf z}\vert^{2n}} \left[L \left(\vert {\bf z} \rangle \langle {\bf z} \vert\right)^{\otimes n}\right] d \text{Re}({\bf z}) d\text{Im}({\bf z}).
\end{equation}
Assuming that $g({\bf z},t)$ becomes negligible as $\vert {\bf z} \vert \to \infty$, integration by parts can be applied to obtain
\begin{equation}\label{eq:gintev}
\int \frac{\partial g({\bf z},t)}{\partial t} \frac{\left(\vert {\bf z} \rangle \langle {\bf z} \vert\right)^{\otimes n}}{\vert{\bf z}\vert^{2n}} d \text{Re}({\bf z}) d\text{Im}({\bf z}) = \int \left[L^\dagger \frac{g({\bf z},t)}{\vert{\bf z}\vert^{2n}}\right]\left(\vert {\bf z} \rangle \langle {\bf z} \vert\right)^{\otimes n} d \text{Re}({\bf z}) d\text{Im}({\bf z}),
\end{equation}
where $L^\dagger$ is the adjoint of $L$, obtained by flipping the signs of terms in $L$ that have odd-order derivatives and reversing the order of the factors in all terms. Next, a choice for the evolution of $g({\bf z},t)$ that is consistent with (\ref{eq:gintev}) is given by equating the integrands:
\begin{equation}\label{eq:gev1}
\frac{\partial g({\bf z},t)}{\partial t} = \vert{\bf z}\vert^{2n} L^\dagger \frac{g({\bf z},t)}{\vert{\bf z}\vert^{2n}}.
\end{equation}

Before deriving explicit evolution equations based on (\ref{eq:gev1}), we make a few remarks about the $f({\bf z}, t)$ and $g({\bf z}, t)$ distributions. Integrating over $\vert {\bf z} \vert = 1$ is preferable since this keeps the quantum states such as $\vert {\bf z} \rangle$ normalized and since points with $\vert {\bf z} \vert \neq 1$ are redundant. The reason for introducing $g({\bf z}, t)$ and considering $\vert {\bf z} \vert \neq 1$ is to handle derivatives. Cartesian derivatives in particular (e.g., $\partial/\partial x_j$) are associated with directions along which $\vert {\bf z} \vert$ generally varies. Still, there is no need to deviate significantly from $\vert {\bf z} \vert = 1$. The distribution $f({\bf z}, t+\Delta t)$ can be derived from $f({\bf z}, t)$ by applying (\ref{eq:extension}), evolving $g({\bf z}, t)$ for $\Delta t$, and applying (\ref{eq:projection}). Using $\Delta t \ll 1$ keeps the distribution close to $\vert {\bf z} \vert = 1$, roughly speaking. It is also possible to project evolution operators onto the $\vert {\bf z} \vert = 1$ surface, which further reduces the need to consider points with $\vert {\bf z} \vert \neq 1$.

Pulling the $\vert{\bf z}\vert^{-2n}$ factor through $L^\dagger$ in (\ref{eq:gev1}) produces
\begin{equation}\label{eq:gev2}
\frac{\partial g({\bf z},t)}{\partial t} = \left(L^\dagger + \vert{\bf z}\vert^{2n} \left[L^\dagger, \vert{\bf z}\vert^{-2n}\right]\right) g({\bf z},t).
\end{equation}
The commutator in (\ref{eq:gev2}) can be evaluated using
\begin{equation}
\left[\bm{\nabla}, \vert{\bf z}\vert^{-2k}\right] = -k {\bf z}^* \vert{\bf z}\vert^{-2(k+1)}.
\end{equation}
For example, with $L$ given by (\ref{eq:L1PI}),
\begin{align}
L^\dagger &= 2 \text{Re}\left(i H_{jk} \nabla_j z_k \right), \\
\vert{\bf z}\vert^{2n} \left[L^\dagger, \vert{\bf z}\vert^{-2n}\right] &= -\frac{2 n}{\vert{\bf z}\vert^2} \text{Re}\left(i H_{jk} z^*_j z_k \right) = 0.
\end{align}
Therefore, the evolution generated by the non-interacting Hamiltonian in (\ref{eq:H1PI}) is reproduced by evolving $g({\bf z},t)$ according to
\begin{equation}\label{eq:g1PI}
\frac{\partial g({\bf z},t)}{\partial t} = 2 \text{Re}\left(i H_{jk} \nabla_j z_k \right) g({\bf z},t).
\end{equation}
Next, the two-particle interaction in (\ref{eq:H2PI}) maps to
\begin{equation}\label{eq:g2PI}
\frac{\partial g({\bf z},t)}{\partial t} = -2 \text{Re}\left( i H_{jklm} \left[ \nabla_j \nabla_k 
- \frac{n}{\vert {\bf z} \vert^2} \left( \nabla_j z^*_k + \nabla_k z^*_j \right) \right] z_l z_m \right) g({\bf z},t),
\end{equation}
and the evolution in (\ref{eq:evX}) maps to
\begin{equation}\label{eq:gX}
\frac{\partial g({\bf z},t)}{\partial t} = \left[2 \text{Im}\left( X_{jk} \nabla_j z_k \right) \right]^2 g({\bf z},t).
\end{equation}
Furthermore, linear combinations of density matrix evolution operators map to the same linear combinations of the $g({\bf z},t)$ evolution operators.

For our purposes, no other evolution terms are needed, so now we conclude this section. Density matrix evolution for systems of $n$ bosonic particles, where each particle has $N$ components, has been mapped to the evolution of a distribution in $2N$-dimensional space. There is also the freedom to project this distribution onto the $(2N-1)$-dimensional hypersphere. If $N$ is large, solving the partial differential equation to evolve this distribution is expensive, but we shall not endeavor to do that. Rather, in the next section, we investigate how the distribution evolution can be related to yet another evolution form: stochastic differential equations.

\section{Connection to stochastic differential equations}\label{sec:consde}

The It\^{o} stochastic differential equation (SDE)
\begin{equation}\label{eq:Ito}
d {\bf r}_t = {\bf F}({\bf r}_t)dt + \sqrt{2 D({\bf r}_t)} d {\bf W}_t,
\end{equation}
where ${\bf r}_t$ is vector of $2N$ real variables indexed by $t$, ${\bf F}({\bf r})$ is a vector function, $D({\bf r})$ is a diffusion matrix, and $d{\bf W}_t$ is a $2N$-dimensional standard Wiener process, has an associated Fokker--Planck equation
\begin{equation}\label{eq:FP}
\frac{\partial f({\bf r}, t)}{\partial t} = \left[-\frac{\partial}{\partial r_j} F_j({\bf r}) + \frac{\partial^2}{\partial r_j \partial r_k} D_{jk}({\bf r}) \right] f({\bf r}, t)
\end{equation}
for the probability distribution $f({\bf r}, t)$ over $2N$-dimensional space. Also, let
\begin{equation}\label{eq:rdef}
{\bf r} \equiv \begin{pmatrix}
{\bf x} \\
{\bf y}
\end{pmatrix} \equiv \begin{pmatrix}
\text{Re}({\bf z}) \\
\text{Im}({\bf z})
\end{pmatrix}.
\end{equation}
Since ${\bf r}$ holds the same information as ${\bf z}$, we shall sometimes use them interchangeably. It is possible for the distribution evolution derived in Sect.~\ref{sec:mapev} to have the form of (\ref{eq:FP}), in which case there is an associated SDE given by (\ref{eq:Ito}). This stochastic system then tracks the density matrix evolution. For instance, each system state is associated with a density matrix of $(\vert {\bf z} \rangle \langle {\bf z} \vert )^{\otimes n}$; and, with $f({\bf z}, t)$ set as the probability distribution of the stochastic system, (\ref{eq:hyp}) gives $\rho(t)$.

Now we check whether specific distribution evolution terms can be related to stochastic evolution terms. First, consider (\ref{eq:g1PI}). Since this is first order in the derivatives, it is associated with deterministic evolution. To determine the particular evolution form, note that
\begin{equation}
\dot{\bf z} = {\bf F}({\bf z}),
\end{equation}
where the time dependence of ${\bf z}$ is implicit, has a Fokker--Planck equation of
\begin{equation}
\frac{\partial f({\bf z}, t)}{\partial t} = -\left(\frac{\partial}{\partial {\bf x}} \cdot \text{Re}\left[{\bf F}({\bf z})\right] + \frac{\partial}{\partial {\bf y}} \cdot \text{Im}\left[{\bf F}({\bf z})\right]\right)f({\bf z}, t),
\end{equation}
and that
\begin{equation}
\left[{\bf z}, \frac{\partial}{\partial {\bf x}} \cdot \text{Re}\left[{\bf F}({\bf z})\right] + \frac{\partial}{\partial {\bf y}} \cdot \text{Im}\left[{\bf F}({\bf z})\right]\right] = -\text{Re}\left[{\bf F}({\bf z})\right] - i \text{Im}\left[{\bf F}({\bf z})\right] = -{\bf F}({\bf z}).
\end{equation}
Therefore,
\begin{equation}
\frac{\partial f({\bf z}, t)}{\partial t} = L f({\bf z}, t),
\end{equation}
where $L$ is linear in spatial derivatives, is the Fokker--Planck equation for deterministic evolution of the form
\begin{equation}\label{eq:extractF}
{\bf F}({\bf z}) = \left[{\bf z}, L\right].
\end{equation}
Applying this with the evolution operator in (\ref{eq:g1PI}) for $L$ produces
\begin{equation}
\begin{split}
F_m({\bf z}) &= \left[z_m, 2 \text{Re}\left(i H_{jk} \nabla_j z_k \right)\right] \\
&= \left[z_m, i H_{jk} \nabla_j z_k - i H^*_{jk} \nabla^*_j z^*_k\right] \\
&= -i H_{mk} z_k.
\end{split}
\end{equation}
So the evolution generated by (\ref{eq:H1PI}) is connected to the simple system of
\begin{equation}\label{eq:ev1PI}
\dot{\bf z} = -i H {\bf z}.
\end{equation}
Since $H = H^\dagger$, this evolution is unitary, so $\vert {\bf z} \vert^2 = \vert {\bf r} \vert^2$ is preserved. In this case there is no need to project the evolution onto $\vert {\bf z} \vert=1$. The ${\bf r}$ evolution takes the form
\begin{equation}\label{eq:rev1}
\dot{\bf r} = \begin{pmatrix}
\text{Re}(-i H) & \text{Im}(i H)\\
\text{Im}(-i H) & \text{Re}(-i H)
\end{pmatrix} {\bf r},
\end{equation}
and ${\bf r} \cdot \dot{\bf r} = 0$ holds because the matrix in (\ref{eq:rev1}) is antisymmetric.

Next, consider the (\ref{eq:gX}) evolution, which has an evolution operator of
\begin{equation}\label{eq:opX}
\left[2 \text{Im}\left( X_{jk} \nabla_j z_k \right) \right]^2 = \left[\frac{\partial}{\partial x_j} \text{Im}\left(X_{jk}z_k\right) - \frac{\partial}{\partial y_j} \text{Re}\left(X_{jk}z_k\right)\right]^2.
\end{equation}
The Fokker--Planck diffusion matrix can be extracted from the evolution operator using commutators; specifically,
\begin{equation}\label{eq:extractD}
\frac12 \left[r_l, \left[r_m, \frac{\partial^2}{\partial r_j \partial r_k} D_{jk}({\bf r})\right]\right] = D_{lm}({\bf r})
\end{equation}
shows that a double commutator can extract $D({\bf r})$ from any evolution operator that is second order in spatial derivatives. Applying this to (\ref{eq:opX}) yields
\begin{equation}\label{eq:XD}
D({\bf z}) = \begin{pmatrix}
\text{Im}\left(X {\bf z}\right) \text{Im}\left(X {\bf z}\right)^T & -\text{Im}\left(X {\bf z}\right) \text{Re}\left(X {\bf z}\right)^T \\
-\text{Re}\left(X {\bf z}\right) \text{Im}\left(X {\bf z}\right)^T & \text{Re}\left(X {\bf z}\right) \text{Re}\left(X {\bf z}\right)^T
\end{pmatrix}
= {\bf u}({\bf z}) {\bf u}({\bf z})^T,
\end{equation}
where
\begin{equation}\label{eq:difu}
{\bf u}({\bf z}) = \begin{pmatrix}
\text{Im}\left(X {\bf z}\right) \\
-\text{Re}\left(X {\bf z}\right)
\end{pmatrix} = \begin{pmatrix}
\text{Im}(X) & \text{Re}(X) \\
-\text{Re}(X) & \text{Im}(X)
\end{pmatrix} {\bf r}.
\end{equation}
This $D({\bf z})$ represents diffusion along ${\bf u}({\bf z})$, which is orthogonal to ${\bf r}$:
\begin{equation}
{\bf r} \cdot {\bf u}({\bf z}) = \text{Re}({\bf z}) \cdot \text{Im}(X {\bf z}) - \text{Im}({\bf z}) \cdot \text{Re}(X {\bf z}) = \text{Im}\left({\bf z}^* \cdot X {\bf z}\right) = 0
\end{equation}
since $X = X^\dagger$. Also, in the same manner that ${\bf r}$ is identified with the complex vector ${\bf z}$, ${\bf u}({\bf z})$ can be identified with the complex vector $-i X {\bf z}$. All of this is similar to the non-interacting evolution case, but the evolution operator in (\ref{eq:opX}) still has more to it. In particular, moving the derivatives to the left produces a commutator term:
\begin{equation}\label{eq:comtm}
\begin{split}
\left[2 \text{Im}\left( X_{jk} \nabla_j z_k \right) \right]^2 - \frac{\partial^2}{\partial r_j \partial r_k} D_{jk}({\bf z}) &= 2 \text{Re}\left(X_{jk} X_{lm} \nabla_j \left[\nabla_l, z_k\right] z_m\right) \\
&= 2 \text{Re}\left[\left(X^2\right)_{jk} \nabla_j z_k\right],
\end{split}
\end{equation}
which is linear in spatial derivatives. Applying (\ref{eq:extractF}) yields the associated deterministic evolution:
\begin{equation}\label{eq:comF}
{\bf F}({\bf z}) = \left[{\bf z}, 2 \text{Re}\left[\left(X^2\right)_{jk} \nabla_j z_k\right]\right] = - X^2 {\bf z}.
\end{equation}
Therefore, (\ref{eq:gX}) is the Fokker--Planck equation for an It\^{o} SDE with ${\bf F}({\bf z})$ and $D({\bf z})$ given by (\ref{eq:comF}) and (\ref{eq:XD}), respectively. Alternatively, this evolution can be formulated as a Stratonovich SDE, in which case the derivatives do not need to be moved to the left and consequently, there is no deterministic evolution term. We choose to still use the It\^{o} formulation while providing the following explanation for the (\ref{eq:comF}) evolution term. In a sense, stochastic evolution is faster than deterministic evolution, with movement scaling as $\sqrt{\Delta t}$ instead of $\Delta t$ for time step $\Delta t$. This speed combined with the fact that the direction ${\bf u}({\bf r})$ is perpendicular to ${\bf r}$ leads to a tendency for $\vert {\bf r} \vert$ to increase, similar to the effect of a centrifugal force. The (\ref{eq:comF}) evolution counters this, such that $\vert {\bf r} \vert^2$ is conserved on average.

Two-particle interaction evolution is both more interesting and more complicated than the previous cases. (\ref{eq:extractF}) and (\ref{eq:extractD}) can be applied to the evolution operator in (\ref{eq:g2PI}) to extract
\begin{align}
{\bf F}({\bf z}) &= \frac{2 n}{\vert {\bf z} \vert^2} B({\bf z}) {\bf z}^*, \label{eq:F2PI}\\
D({\bf z}) &= \frac12 \begin{pmatrix}
\text{Re}\left[B({\bf z})\right] & \text{Im}\left[B({\bf z})\right] \\
\text{Im}\left[B({\bf z})\right] & -\text{Re}\left[B({\bf z})\right]
\end{pmatrix}\label{eq:D2PI},
\end{align}
where $B({\bf z})$ is an $N \times N$ matrix with entries given by
\begin{equation}
B_{jk}({\bf z}) = - \frac{i}{2} \left(H_{jklm} + H_{kjlm}\right) z_l z_m.
\end{equation}
The deterministic part of the evolution [given by (\ref{eq:F2PI})] preserves $\vert {\bf z} \vert$ since
\begin{equation}
\begin{split}
\text{Re}\left({\bf z}^* \cdot {\bf F}({\bf z})\right) = \frac{2n}{\vert {\bf z} \vert^2} \text{Re}\left(-i H_{jklm} z^*_j z^*_k z_l z_m \right) = 0
\end{split}
\end{equation}
with the second equality holding due to $\hat{H} = \hat{H}^\dagger$ for the $\hat{H}$ in (\ref{eq:H2PI}). So if the evolution was just $\dot{\bf z} = {\bf F}({\bf z})$, then the trajectories starting with $\vert {\bf z} \vert=1$ would remain confined to the $\vert {\bf z} \vert=1$ hypersphere. Furthermore, since a factor of the particle count $n$ shows up in (\ref{eq:F2PI}) but not in (\ref{eq:D2PI}), there is a limit that can be taken to realize this scenario. Replacing (\ref{eq:H2PI}) with
\begin{equation}\label{eq:intH}
\hat{H} = \frac{1}{2n} H_{jklm} \hat{a}_j^\dagger \hat{a}^\dagger_k \hat{a}_l \hat{a}_m
\end{equation}
and taking the $n \to \infty$ limit results in ${\bf F}({\bf z}) \to B({\bf z}) {\bf z}^* / \vert {\bf z} \vert^2$ and $D({\bf z}) \to 0$. This leaves deterministic nonlinear evolution corresponding to a bosonic mean-field theory. Then, for finite $n$, the diffusive terms can be viewed as finite-$n$ corrections to this mean-field theory approximation. However, this picture is complicated by the fact that the $D({\bf z})$ in (\ref{eq:D2PI}) generally has eigenvalues of both signs, since it is symmetric and traceless.

Writing the distribution evolution in the form of (\ref{eq:FP}) is not enough to ensure a correspondence with an SDE because  (\ref{eq:Ito}) is not valid unless the diffusion matrix is positive semi-definite. This is related to the irreversibility of stochastic processes. Of course, unitary quantum evolution is reversible, so it is not surprising that (\ref{eq:D2PI}) fails to have this property. On the other hand, the open quantum system evolution of (\ref{eq:evX}) is irreversible, and its associated diffusion matrix [given by (\ref{eq:XD})] is positive semi-definite. Now consider open quantum systems that have both unitary interactions and evolution of the (\ref{eq:evX}) form. The diffusion matrices for the various evolution terms combine linearly, which may result in the elimination of negative (i.e., reverse) diffusion. In such a scenario, the interactions with the environment are strong enough that an initially separable $\rho$ stays separable; in other words, decoherence prevents entanglement from developing.

Keeping $\rho$ separable does not necessarily cause the quantum dynamics to be trivial. First, it is important to recognize that a lack of entanglement between $N$-state bosonic particles is different from a lack of entanglement between qubits. We are particularly interested in cases with $N \gg 1$, such that the stochastic system is also very large. Yet, one can in principle simulate arbitrary quantum dynamics on a single bosonic particle with sufficiently large $N$, and then the entanglement between particles is irrelevant. This is connected to the fact that representing the $N$ components of one particle with $\Theta(\log N)$ qubits generically requires significant entanglement between those qubits. Additionally, even when there is not much entanglement between qubits, mixed state dynamics can remain interesting. The proof by Vidal \cite{Vidal2003} that quantum systems with low entanglement can be efficiently simulated classically is only for pure states, whereas open quantum systems with very little entanglement still appear to be computationally powerful \cite{Datta2005, OneCleanQubit2014}.

\subsection{Elimination of negative eigenvalues}
Now we show that, for any two particle interaction in the form of (\ref{eq:H2PI}), it is possible to add evolution terms of the form in (\ref{eq:evX}) to obtain an open quantum system that is associated with an SDE. The density matrix evolution
\begin{equation}
\begin{split}\label{eq:phoev}
\dot{\rho} &= -i \left[\hat{H}, \rho\right] - \sum_m \left[ \hat{X}_m, \left[ \hat{X}_m, \rho \right]\right], \\
\hat{H} &= H^0_{jk} \hat{a}^\dagger_j \hat{a}_k + \frac{1}{2n} H_{jklm} \hat{a}_j^\dagger \hat{a}^\dagger_k \hat{a}_l \hat{a}_m, \\
\hat{X}_m &= \frac{1}{\sqrt{n}} X_{mjk} \hat{a}^\dagger_j \hat{a}_k,
\end{split}
\end{equation}
where all operators are Hermitian, maps to distribution evolution in the form of (\ref{eq:FP}) with
\begin{align}
{\bf F}({\bf z}) &= -i H^0 {\bf z} + \frac{B({\bf z}) {\bf z}^*}{\vert {\bf z} \vert^2} - \frac{1}{n} \sum_m X_m^2 {\bf z}, \label{eq:fullF}\\
D({\bf z}) &= \frac{1}{n}\left[\frac14 \begin{pmatrix}
\text{Re}\left[B({\bf z})\right] & \text{Im}\left[B({\bf z})\right] \\
\text{Im}\left[B({\bf z})\right] & -\text{Re}\left[B({\bf z})\right]
\end{pmatrix} + \sum_m {\bf u}_m({\bf z}) {\bf u}_m({\bf z})^T\right], \label{eq:Dfull}\\
{\bf u}_m({\bf z}) &= \begin{pmatrix}
\text{Im}\left(X_m {\bf z}\right) \\
-\text{Re}\left(X_m {\bf z}\right)
\end{pmatrix}, \\
B_{jk}({\bf z}) &= - \frac{i}{2} \left(H_{jklm} + H_{kjlm}\right) z_l z_m, \\
\left(X_m\right)_{jk} &= X_{mjk} \label{eq:lastX}.
\end{align}
Ideally, we would find $X_m$ to make $D({\bf r})$ [$\equiv D({\bf z})$] positive semi-definite, but there is an issue that must be addressed first. Unlike (\ref{eq:XD}), the eigenvectors of (\ref{eq:D2PI}) and (\ref{eq:Dfull}) are not generally orthogonal to ${\bf r}$. However, since the distribution can always be projected onto $\vert {\bf r} \vert = 1$ using (\ref{eq:projection}), the components along ${\bf r}$ are of no consequence. In particular,
\begin{equation}
\int_0^\infty \left[\left(c\frac{\partial}{\partial r} + \frac{\partial}{\partial r_{\perp}}\right)^2 h(r)\right] dr = \int_0^\infty \left[\left(\frac{\partial}{\partial r_{\perp}}\right)^2 h(r)\right] dr,
\end{equation}
where $r$ is the direction along ${\bf r}$, $r_{\perp}$ is any direction perpendicular to ${\bf r}$, and $h(r)$ is any distribution that becomes negligible for $r \to 0$ and $r \to \infty$. Therefore, each eigenvector of $D({\bf r})$ can be replaced with its projection onto the hypersphere, which is equivalent to subtracting off its component along ${\bf r}$. Applying this to all eigenvectors results in the replacement of $D({\bf r})$ with
\begin{equation}\label{eq:Dperp}
D_{\perp}({\bf r}) \equiv D({\bf r}) - {\bf r} {\bf r}^T D({\bf r}) - D({\bf r}) {\bf r} {\bf r}^T - \left[{\bf r} \cdot D({\bf r}) {\bf r}\right] {\bf r} {\bf r}^T.
\end{equation}
Replacing $D({\bf r})$ with $D_{\perp}({\bf r})$ is crucial because the positive diffusion from (\ref{eq:XD}) is always perpendicular to ${\bf r}$, while the negative diffusion from (\ref{eq:D2PI}) can be partially along ${\bf r}$. Consequently, it is not generally possible to find $X_m$ that make (\ref{eq:Dfull}) positive semi-definite. Yet, it is possible to find $X_m$ that make $D_{\perp}({\bf r})$ positive semi-definite, as the following argument shows. Any real vector ${\bf s}$ can be written as
\begin{equation}\label{eq:sw}
{\bf s} = \begin{pmatrix}
\text{Re}({\bf w}) \\
\text{Im}({\bf w})
\end{pmatrix},
\end{equation}
and its dot product with ${\bf r}$ is
\begin{equation}
{\bf s} \cdot {\bf r} = \text{Re}({\bf w}) \cdot \text{Re}({\bf z}) + \text{Im}({\bf w}) \cdot \text{Im}({\bf z}) = \text{Re}\left({\bf w}^* \cdot {\bf z}\right).
\end{equation}
Next, the matrix
\begin{equation}\label{eq:Xsp}
X = i {\bf w} {\bf z}^\dagger - i {\bf z} {\bf w}^\dagger + i ({\bf w}^* \cdot {\bf z}) {\bf z} {\bf z}^\dagger
\end{equation}
is Hermitian provided that $\text{Re}\left({\bf w}^* \cdot {\bf z}\right) = {\bf s} \cdot {\bf r} = 0$, and it satisfies
\begin{equation}
-i X {\bf z} = {\bf w}
\end{equation}
for normalized ${\bf z}$. Therefore, using (\ref{eq:Xsp}) for the $X$ in (\ref{eq:XD}) gives diffusion along ${\bf s}$, where ${\bf s}$ can be any vector that is orthogonal to ${\bf r}$. In particular, ${\bf s}$ can be any $D_{\perp}({\bf r})$ eigenvector with a negative eigenvalue, and $X$ can be scaled to cancel that $D_{\perp}({\bf r})$ component when added as one of the $X_m$ in (\ref{eq:phoev})--(\ref{eq:lastX}). Repeating this for each negative eigenvalue will cause $D_{\perp}({\bf r})$ to become positive semi-definite.

So the negative eigenvalues in $D_{\perp}({\bf r})$ can be eliminated for a particular ${\bf r}$, but $D_{\perp}({\bf r})$ needs to be positive semi-definite for every normalized ${\bf r}$ to guarantee the validity of the associated SDE, which is (\ref{eq:Ito}) with $D_{\perp}({\bf r})$ in place of $D({\bf r})$. As ${\bf r}$ varies, both the eigenvectors of $D_{\perp}({\bf r})$ and the diffusion direction [given by (\ref{eq:difu})] from a fixed $X$ vary, such that the cancelation of negative diffusion at one ${\bf r}$ does not generally transfer over to other ${\bf r}$. Yet it is possible to select any orthonormal set of ${\bf s}$ vectors of size $2N-1$, all orthogonal to ${\bf r}$, and apply (\ref{eq:sw}) and (\ref{eq:Xsp}) to each to get a corresponding set of $X$ matrices. When these are added to the $X_m$ set, the resulting change to (\ref{eq:Dfull}) at this particular ${\bf r}$ is $(I - {\bf r} {\bf r}^T)/n$, where $I$ is the $2N \times 2N$ identity matrix, and the same holds for (\ref{eq:Dperp}). Furthermore, near this ${\bf r}$, the change is approximately the same; in particular, positive diffusion is still added in every direction besides the radial one. Then, by repeating these $X_m$ additions for a finite number of ${\bf r}$ vectors, positive diffusion can be added in all directions over the hypersphere, which allows for making $D_{\perp}({\bf r})$ positive semi-definite at all ${\bf r}$.

As a simple demonstration, consider $N=2$ and $X_m$ given by the Pauli matrices:
\begin{align}
X_0 &= \sigma_0 = \begin{pmatrix}
0 & 1 \\
1 & 0
\end{pmatrix},&
X_1 &= \sigma_1 = \begin{pmatrix}
0 & -i \\
i & 0
\end{pmatrix},&
X_2 &= \sigma_2 = \begin{pmatrix}
1 & 0 \\
0 & -1
\end{pmatrix}.
\end{align}
The diffusion vectors ${\bf u}_m({\bf r})$ can be expressed as
\begin{equation}
{\bf u}_m({\bf r}) = A_m {\bf r},\\
\end{equation}
where
\begin{equation}
A_m \equiv \begin{pmatrix}
\text{Im}(X_m) & \text{Re}(X_m) \\
-\text{Re}(X_m) & \text{Im}(X_m)
\end{pmatrix}. \label{eq:Am}
\end{equation}
The $A_m$ matrices are always anti-symmetric, but in this case, they have even more structure:
\begin{equation}\label{eq:antirel}
\left\{A_j, A_k\right\} = -2\delta_{jk}.
\end{equation}
Due to this structure, the ${\bf u}_m({\bf r})$ vectors are orthonormal:
\begin{equation}
\begin{split}
{\bf u}_j({\bf r}) \cdot {\bf u}_k({\bf r}) &= \frac12 \left[ {\bf u}_j({\bf r}) \cdot {\bf u}_k({\bf r}) + {\bf u}_k({\bf r}) \cdot {\bf u}_j({\bf r})\right] \\
&= \frac12 {\bf r} \cdot \left(A_j^T A_k + A_k^T A_j \right) {\bf r} \\
&= -\frac12 {\bf r} \cdot \left\{A_j, A_k\right\} {\bf r} \\
&= \delta_{jk}.
\end{split}
\end{equation}
It follows that, with $\hat{H} = 0$ in (\ref{eq:phoev}), these $X_m$ yield
\begin{equation}
D({\bf r}) = D_\perp({\bf r}) = \frac{1}{n} \left( I - {\bf r} {\bf r}^T \right).
\end{equation}
Canceling all negative diffusion in the $N=2$ case is thus straightforward: it suffices to choose $X_m = \sqrt{\gamma} \sigma_m$ for $m \in \{0,1,2\}$, where
\begin{equation}
\begin{split}
\gamma &= \max_{\vert{\bf r}\vert=1} \left\Vert C_\perp({\bf r}) \right\Vert,\\
C_\perp({\bf r}) &= C({\bf r}) - {\bf r} {\bf r}^T C({\bf r}) - C({\bf r}) {\bf r} {\bf r}^T - \left[{\bf r} \cdot C({\bf r}) {\bf r}\right] {\bf r} {\bf r}^T, \\
C({\bf r}) &= \frac14 \begin{pmatrix}
\text{Re}\left[B({\bf r})\right] & \text{Im}\left[B({\bf r})\right] \\
\text{Im}\left[B({\bf r})\right] & -\text{Re}\left[B({\bf r})\right]
\end{pmatrix},
\end{split}
\end{equation}
$\Vert \cdot \Vert$ denotes the spectral norm, and $B({\bf r})$ is the same matrix as $B({\bf z})$ in (\ref{eq:fullF}).

Unfortunately, this simple solution for $X_m$ with $N=2$ does not generalize. The size of the largest set of anticommuting $2^q p \times 2^q p$ matrices, where $p$ is odd, is $2q+1$ \cite{Newman1932}. So it is not generally possible to find $2N-1$ matrices of size $2N \times 2N$ that satisfy (\ref{eq:antirel}), even before restricting them to have the form of (\ref{eq:Am}). The argument given previously still ensures that some number $M(2N-1)$ of $X_m$ matrices is adequate to make $D_\perp$ positive semi-definite, where $M$ is the number of ${\bf r}$ points used, but we choose not to construct explicit solutions based on that argument. If too much positive diffusion is added, the dynamics due to the two-particle interactions can get washed out by noise. Therefore, we think it is better to seek $X_m$ matrices that are effective at canceling negative diffusion for specific two-particle interactions of interest. Also, eliminating all negative diffusion is not strictly necessary: if the remaining negative eigenvalues of $D_\perp({\bf z})$ are small, then the effect of neglecting them is also small, which is made precise in the following subsection.

\subsection{Error bound for approximate stochastic evolution}\label{sec:apx}

The absolute sum of the negative eigenvalues of any symmetric $S$ is given by
\begin{equation}\label{eq:nevs}
\left\Vert S \right\Vert_* - \text{Tr}(S),
\end{equation}
where $\left\Vert \cdot \right\Vert_*$ denotes the trace norm. For $D({\bf z})$ and $D_\perp({\bf z})$ in particular,
\begin{align}
\alpha({\bf z}) &\equiv \left\Vert D({\bf z}) \right\Vert_* - \text{Tr}\left[D({\bf z})\right],\\
\alpha_\perp({\bf z}) &\equiv \left\Vert D_\perp({\bf z}) \right\Vert_* - \text{Tr}\left[D_\perp({\bf z})\right]
\end{align}
are related by
\begin{equation}\label{eq:nerel}
\alpha_\perp({\bf z}) \leq \alpha({\bf z}),
\end{equation}
which can be seen as follows. For a single eigencomponent of $D({\bf z})$, projection produces a new eigencomponent with absolute eigenvalue no greater than before, thus never increasing the associated (\ref{eq:nevs}) value. Furthermore, (\ref{eq:nevs}) is subadditive, so summing the projected eigencomponents yields (\ref{eq:nerel}).

Next, consider the effect of dropping all negative eigencomponents of $D_\perp({\bf z})$. This means that $D_\perp({\bf z})$ is replaced by
\begin{equation}\label{eq:Dplus}
D^+_\perp({\bf z}) \equiv \frac12 \left(D_\perp({\bf z}) + \sqrt{D^2_\perp({\bf z})}\right),
\end{equation}
which is always positive semi-definite and differs from $D_\perp({\bf z})$ by
\begin{equation}
\left\Vert D^+_\perp({\bf z}) - D_\perp({\bf z}) \right\Vert_* = \alpha_\perp({\bf z}).
\end{equation}
The stochastic system given by (\ref{eq:Ito}) with $D^+_\perp({\bf r})$ in place of $D({\bf r})$ is associated with density matrix evolution, which we denote $\rho^+(t)$, obtained by inputting the SDE probability distribution $f({\bf z}, t)$ into (\ref{eq:hyp}). The density matrix evolution given by (\ref{eq:phoev}) differs from the $\rho^+(t)$ evolution based on the difference between $D^+_\perp({\bf z})$ and $D_\perp({\bf z})$. Treating this as an error, the trace preservation and positivity properties of density matrix evolution [including (\ref{eq:phoev})] imply a simple error propagation bound:
\begin{equation}
\left\Vert \rho^+(t) - \rho(t) \right\Vert_* \leq \int_0^t \left\Vert \dot{\rho}^+(t') - \dot{\rho}(t')\right\Vert_* dt'.
\end{equation}
Moreover, because $\rho^+(t)$ is represented as (\ref{eq:hyp}) with $f({\bf z}, t)$ a probability distribution,
\begin{equation}
\left\Vert \dot{\rho}^+(t) - \dot{\rho}(t)\right\Vert_* \leq \max_{\vert {\bf z} \vert=1} \left\Vert \Delta({\bf z}) \right\Vert_*,
\end{equation}
where $\Delta({\bf z}) = \dot{\rho}^+ - \dot{\rho}$ for $\rho = (\vert {\bf z} \rangle \langle {\bf z} \vert)^{\otimes n}$. This can be further decomposed by summing the contributions to $\Delta({\bf z})$ from each eigencomponent of ${D^+_\perp({\bf z}) - D_\perp({\bf z})}$. Combining these decompositions yields
\begin{equation}\label{eq:drhob}
\left\Vert \rho^+(t) - \rho(t) \right\Vert_* \leq \beta(n) t \max_{\vert {\bf z} \vert=1} \alpha_\perp({\bf z}),
\end{equation}
where $\beta(n)$ is the maximum trace norm of the change in $\dot{\rho}$ for any $\rho = (\vert {\bf z} \rangle \langle {\bf z} \vert)^{\otimes n}$ that is caused by a change of
\begin{equation}
{\bf s} {\bf s}^T
\end{equation}
to $D({\bf z})$ for any ${\bf s}$ that is normalized and orthogonal to ${\bf r}$. To bound $\beta(n)$, we can use the previous result that diffusion along a direction ${\bf s}$ is mapped to by the evolution in (\ref{eq:evX}) with $\hat{X}$ given by (\ref{eq:formX}), (\ref{eq:sw}), and (\ref{eq:Xsp}). Applying that mapping in the other direction and canceling the commutator term [cf. (\ref{eq:comtm})] leads to
\begin{equation}\label{eq:drhoz}
\begin{split}
\beta(n) &=\begin{multlined}[t] \max_{\vert{\bf z}\vert=1, \vert{\bf w}\vert=1, \text{Re}({\bf z}^* \cdot {\bf w})=0} \Big\Vert \Big( \left[\hat{X}({\bf z},{\bf w})^2 - \hat{Y}({\bf z},{\bf w})\right] (\vert {\bf z} \rangle \langle {\bf z} \vert)^{\otimes n} \\
- \hat{X}({\bf z},{\bf w}) (\vert {\bf z} \rangle \langle {\bf z} \vert)^{\otimes n} \hat{X}({\bf z},{\bf w}) \Big) + \text{h.c.} \Big\Vert_*, \end{multlined}\\
\hat{X} &= X({\bf z},{\bf w})_{jk} \hat{a}^\dagger_j \hat{a}_k,\\
\hat{Y} &= \left[X({\bf z},{\bf w})^2\right]_{jk} \hat{a}^\dagger_j \hat{a}_k,
\end{split}
\end{equation}
where $X({\bf z},{\bf w})$ is given by (\ref{eq:Xsp}). Through straightforward manipulations, (\ref{eq:drhoz}) can be rewritten as
\begin{equation}
\begin{split}
\beta(n) &=\begin{multlined}[t] \max_{\vert{\bf z}\vert=1, \vert{\bf w}\vert=1, \text{Re}({\bf z}^* \cdot {\bf w})=0} \Big\Vert \Big(\left[\hat{M} - 2i\left({\bf z}^* \cdot {\bf w}\right)\right] \hat{M} (\vert {\bf z} \rangle \langle {\bf z} \vert)^{\otimes n} \\
- \hat{M} (\vert {\bf z} \rangle \langle {\bf z} \vert)^{\otimes n} \hat{M}^\dagger + n \left({\bf z}^* \cdot {\bf w}\right)^2 (\vert {\bf z} \rangle \langle {\bf z} \vert)^{\otimes n}\Big) + \text{h.c.}\Big\Vert_*, \end{multlined}\label{eq:rhod2}\\
\hat{M} &= i \left[{\bf w} - \left({\bf z}^* \cdot {\bf w}\right) {\bf z}\right] \cdot \hat{\bf a}^\dagger \left({\bf z}^* \cdot \hat{\bf a}\right).
\end{split}
\end{equation}
The $\hat{M}$ operator can also be written $\hat{b}^\dagger \hat{z}$, where $\hat{z} = {\bf z}^* \cdot \hat{\bf a}$ annihilates a ${\bf z}$ particle, $\hat{b}^\dagger = {\bf b} \cdot \hat{\bf a}^\dagger$ creates a ${\bf b}$ particle, and ${\bf b}$ is orthogonal to ${\bf z}$. Next, $\beta(n)$ can be bounded by applying
\begin{equation}
\left\Vert \vert \varphi \rangle \langle \theta \vert \right\Vert_* = \sqrt{\langle \varphi \vert \varphi \rangle \langle \theta \vert \theta \rangle},
\end{equation}
which holds for arbitrary $\vert \varphi \rangle$ and $\vert \theta \rangle$, to the terms in (\ref{eq:rhod2}). This produces
\begin{equation}\label{eq:betabd}
\begin{split}
\beta(n) &\leq\begin{multlined}[t] \max_{\vert{\bf z}\vert=1, \vert{\bf w}\vert=1, \text{Re}({\bf z}^* \cdot {\bf w})=0} 2\Big[\sqrt{2n(n-1)}\vert {\bf w} - \left({\bf z}^* \cdot {\bf w}\right) {\bf z} \vert^2 + n \\
+ 2 \sqrt{n} \vert{\bf z}^* \cdot {\bf w}\vert \left\vert {\bf w} - \left({\bf z}^* \cdot {\bf w}\right) {\bf z} \right\vert \Big]\end{multlined} \\
&\leq 6 n,
\end{split}
\end{equation}
where the coefficient of 6 in the second inequality is just a simple, somewhat arbitrary choice. The important point is that $\beta(n) = \mathcal{O}(n)$, which is better than the na\"{i}ve scaling of $\mathcal{O}(n^2)$ based on how the $\hat{X}$ operator in (\ref{eq:drhoz}) scales as $\mathcal{O}(n)$. Applying (\ref{eq:betabd}) to (\ref{eq:drhob}) yields the bound
\begin{equation}\label{eq:frhod}
\left\Vert \rho^+(t) - \rho(t) \right\Vert_* \leq 6 n t \max_{\vert {\bf z} \vert=1} \alpha_\perp({\bf z})
\end{equation}
for the difference in the density matrices between the unmodified quantum system and the stochastic system given by (\ref{eq:Ito}) with $D^+_\perp({\bf r})$ in place of $D({\bf r})$.

\section{Implementation strategies and efficiency}\label{eq:stqa}

\subsection{Stochastic quantum approach}
The correspondence between the SDE of
\begin{equation}\label{eq:SDE}
d {\bf r}_t = {\bf F}({\bf r}_t)dt + \sqrt{2 D_\perp({\bf r}_t)} d {\bf W}_t
\end{equation}
and the open quantum system of (\ref{eq:phoev}) with $\hat{X}_m$ selected to ensure positive semi-definite $D_\perp({\bf r})$, as discussed in Sect.~\ref{sec:consde}, shows that a quantum system can simulate (\ref{eq:SDE}). Additionally, for large $n$, the objects in (\ref{eq:fullF}) and (\ref{eq:Dfull}) that scale as $\mathcal{O}(1/n)$ are small, such that the SDE approximates the deterministic system of
\begin{equation}\label{eq:mfev}
\dot{\bf z} = -i H^0 {\bf z} + \frac{B({\bf z}) {\bf z}^*}{\vert {\bf z} \vert^2}.
\end{equation}
Of particular interest is the possibility for a quantum algorithm to approximate some of these systems with complexity polynomial in $\log N$ and $t$. Each $N$-state bosonic particle can be represented with $\mathcal{O}(\log N)$ qubits, but determining when the (\ref{eq:phoev}) evolution can be simulated with $\mathcal{O}[\text{poly}(\log N)]$ complexity is more involved.

Conventional quantum computers are based on unitary dynamics, so the open quantum evolution of (\ref{eq:phoev}) should first be represented unitarily. One option is to construct a larger, unitary quantum system such that (\ref{eq:phoev}) describes the dynamics of a subsystem of this system. The correspondence between the (\ref{eq:evX}) evolution and unconditional continuous measurement processes \cite{ContinuousMeasurements1987, ContinuousMeasurements2020} could be used to construct this unitary quantum system. But instead, we consider a more direct approach. Let
\begin{equation}
\begin{split}\label{eq:psisde0}
\vert \psi[q(t)] \rangle &= e^{-i q(t) \hat{Y}} \vert \psi(0) \rangle, \\
dq(t) &= dW(t),\\
q(0) &= 0,
\end{split}
\end{equation}
where $\hat{Y}$ is Hermitian, and $dW(t)$ is a standard Wiener process. The $\vert \psi(t) \rangle$ state evolves stochastically, but its normalization is preserved. Further analysis is facilitated by applying the well-known correspondence between random walks and Wiener processes. Specifically, $dq(t) = dW(t)$ evolution is recovered from the $\Delta t \to 0$ limit of a random walk with steps
\begin{equation}
q(t + \Delta t) = q(t) \pm \sqrt{\Delta t},
\end{equation}
where the $\pm$ branches occur with equal probability. Next, when a random walk step is applied to (\ref{eq:psisde0}), the density matrix can hold the classical probability distribution over the two branches:
\begin{equation}
\begin{split}\label{eq:branches}
\rho(t + \Delta t) &= \frac12 \sum_{\pm} e^{\mp i \hat{Y} \sqrt{\Delta t}} \rho(t) e^{\pm i \hat{Y} \sqrt{\Delta t}} \\
&= \rho(t) + \frac12 \left(2\hat{Y} \rho(t) \hat{Y} - \hat{Y}^2\rho(t) - \rho(t) \hat{Y}^2\right) \Delta t + \mathcal{O}\left[(\Delta t)^2\right].
\end{split}
\end{equation}
Therefore, the density matrix evolution of
\begin{equation}\label{eq:rhoH}
\dot{\rho} = -\frac12 [\hat{Y}, [\hat{Y}, \rho]]
\end{equation}
results in
\begin{equation}
\rho(t) = \int f(q, t) \vert \psi(q) \rangle \langle \psi(q) \vert dq,
\end{equation}
where $f(q, t)$ is the probability distribution of the quantum states. This means that evolution terms of the form in (\ref{eq:rhoH}) can be reproduced using stochastic quantum state evolution of the form in (\ref{eq:psisde0}). In particular, the $\rho$ evolution in (\ref{eq:phoev}) with $\rho(0) = \vert \psi(0) \rangle \langle \psi(0) \vert$ corresponds to stochastic quantum evolution of
\begin{equation}
\begin{split}\label{eq:psisde}
\vert \psi\left[t, {\bf q}(t)\right] \rangle &= \exp\left[-i \left(\hat{H} t + \sqrt{2} \sum_m q_m(t) \hat{X}_m\right)\right] \vert \psi(0) \rangle, \\
d{\bf q}(t) &= d{\bf W}(t),\\
{\bf q}(0) &= {\bf 0}.
\end{split}
\end{equation}
It is worth mentioning that combining evolution terms is more subtle when those terms are stochastic. The basic result of
\begin{equation}\label{eq:splitting}
e^{-i \hat{H}_1 \Delta t} e^{-i \hat{H}_2 \Delta t} + \mathcal{O}\left[ (\Delta t)^2\right] = e^{-i(\hat{H}_1+\hat{H}_2)\Delta t} = e^{-i \hat{H}_1 \Delta t} e^{-i \hat{H}_2 \Delta t} + \mathcal{O}\left[ (\Delta t)^2\right]
\end{equation}
allows for simply combining deterministic evolution terms because the $\mathcal{O}[ (\Delta t)^2]$ errors, accumulated over $t/\Delta t$ steps, still vanish for $\Delta t \to 0$. But stochastic random walk steps have $\sqrt{\Delta t}$ in place of $\Delta t$ in (\ref{eq:splitting}), which breaks this property. However, this complication turns out to be unimportant, which can be seen as follows. The extension of (\ref{eq:branches}) to a pair of independent stochastic terms satisfies
\begin{equation}
\begin{split}\label{eq:rhosplit}
\rho(t + \Delta t) &= \frac14 \sum_{s_0 = \pm} \sum_{s_1 = \pm} e^{- i \left( s_0 \hat{Y}_0 + s_1 \hat{Y}_1\right)\sqrt{\Delta t}} \rho(t) e^{ i \left( s_0 \hat{Y}_0 + s_1 \hat{Y}_1\right)\sqrt{\Delta t}} \\
&= \frac14 \sum_{s_0 = \pm} \sum_{s_1 = \pm} e^{-i s_0 \hat{Y}_0 \sqrt{\Delta t}} e^{-i s_1 \hat{Y}_1 \sqrt{\Delta t}} \rho(t) e^{i s_1 \hat{Y}_1 \sqrt{\Delta t}} e^{i s_0 \hat{Y}_0 \sqrt{\Delta t}} + \mathcal{O}\left[(\Delta t)^2\right] \\
&= \frac14 \sum_{s_0 = \pm} \sum_{s_1 = \pm} e^{-i s_1 \hat{Y}_1 \sqrt{\Delta t}} e^{-i s_0 \hat{Y}_0 \sqrt{\Delta t}} \rho(t) e^{i s_0 \hat{Y}_0 \sqrt{\Delta t}} e^{i s_1 \hat{Y}_1 \sqrt{\Delta t}} + \mathcal{O}\left[(\Delta t)^2\right] \\
&= \rho(t) + \frac12 \sum_{k=0}^1 \left(2\hat{Y}_k \rho(t) \hat{Y}_k - \hat{Y}_k^2\rho(t) - \rho(t) \hat{Y}_k^2\right) \Delta t + \mathcal{O}\left[(\Delta t)^2\right],
\end{split}
\end{equation}
which shows that there is a freedom to rearrange the stochastic evolution terms, analogous to (\ref{eq:splitting}). This freedom arises because the sum over branches causes the cancelation of all terms that have an odd number of appearances of any $\hat{Y}_k$, which is just a consequence of the random walk step distribution having zero mean. One implication is that, when the (\ref{eq:psisde}) evolution is implemented using finite time steps, there is the usual freedom to split up the evolution into parts. For example, applying any one of
\begin{equation}\label{eq:splitx}
\begin{split}
\exp\left[-i \left(\hat{H} \Delta t + \sqrt{2} \sum_m \left[q_m(t+\Delta t)-q_m(t)\right] \hat{X}_m\right)\right], \\
\exp\left(-i \hat{H} \Delta t\right)\exp\left(-i \sqrt{2} \sum_m \left[q_m(t+\Delta t)-q_m(t)\right] \hat{X}_m\right), \\
\left[\prod_m \exp\left(-i \sqrt{2} \left[q_m(t+\Delta t)-q_m(t)\right] \hat{X}_m\right)\right] \exp\left(-i \hat{H} \Delta t\right)
\end{split}
\end{equation}
to $\vert \psi(t) \rangle$ results in the appropriate $\rho(t + \Delta t) - \rho(t)$ with $\mathcal{O}\left[(\Delta t)^2\right]$ error, even though there may be $\Theta(\Delta t)$ differences between the resulting $\vert \psi(t+\Delta t) \rangle$.

\subsection{Output quantities}
After the stochastic quantum evolution is performed, some useful output should be computed. The expectation value $O(t)$ of an observable $\hat{O}$,
\begin{equation}\label{eq:observe}
O(t) = \text{Tr}\left[\hat{O} \rho(t)\right],
\end{equation}
can be evaluated on (\ref{eq:hyp}) to give
\begin{equation}
O(t) = \int_{\vert{\bf z}\vert=1} f({\bf z},t) \text{Tr}\left[\hat{O} \left(\vert {\bf z} \rangle \langle {\bf z} \vert\right)^{\otimes n}\right] d \text{Re}({\bf z}) d\text{Im}({\bf z}).
\end{equation}
So, when $f({\bf z},t)$ is a probability distribution, $O(t)$ is the expectation value of a quantity
\begin{equation}
y({\bf z}) \equiv \text{Tr}\left[\hat{O} \left(\vert {\bf z} \rangle \langle {\bf z} \vert\right)^{\otimes n}\right]
\end{equation}
that depends on the state ${\bf z}$ of the stochastic system at the final time $t$. A variety of polynomials of the components of ${\bf z}$ and ${\bf z}^*$ can be achieved for $y({\bf z})$ by selecting an appropriate $\hat{O}$. For example,
\begin{equation}
\hat{O} = \frac{1}{n} O_{jk} \hat{a}^\dagger_j \hat{a}_k
\end{equation}
produces
\begin{equation}
y({\bf z}) = O_{jk} z^*_j z_k.
\end{equation}
Meanwhile, in terms of the probability distribution $f({\bf q}, t)$ of the stochastic quantum system, the density matrix is
\begin{equation}\label{eq:sqrho}
\rho(t) = \int f({\bf q}, t) \vert \psi(t, {\bf q}) \rangle \langle \psi(t, {\bf q}) \vert d{\bf q},
\end{equation}
and (\ref{eq:observe}) evaluates to
\begin{equation}
O(t) = \int f({\bf q}, t) \text{Tr}\left[\hat{O} \vert \psi(t, {\bf q}) \rangle \langle \psi(t, {\bf q}) \vert \right]d{\bf q}.
\end{equation}
Estimation of $O(t)$ can therefore proceed through the usual process of averaging the measurements of the observable $\hat{O}$ over many runs, although now there is randomness from both the stochastic evolution and the final measurement. Another consideration is that, in any quantum algorithm, approximations will be made. So the achieved density matrix $\tilde{\rho}(t)$ will generally differ from the ideal $\rho(t)$, but the impact that this will have on the output can be bounded. In particular, for an exact output $O(t)$ and an approximation $\tilde{O}(t)$ given by
\begin{equation}
\tilde{O}(t) \equiv \text{Tr}\left[\hat{O} \tilde{\rho}(t)\right],
\end{equation}
the bound
\begin{equation}\label{eq:outdb}
\left\vert \tilde{O}(t) - O(t)\right\vert \leq \left\Vert \hat{O} \right\Vert \left\Vert \tilde{\rho}(t) - \rho(t) \right\Vert_*
\end{equation}
follows from the H\"{o}lder inequality for Schatten norms [(1.174) in \cite{TQIbook}].

\subsection{Stochastic discrete nonlinear Schr\"{o}dinger equation}\label{sec:DNSE}

We now consider a specific system of interest. Let
\begin{equation}\label{eq:HDNLS}
H_{jklm} = \delta_{jk} \delta_{kl} \delta_{lm}.
\end{equation}
Then the components of (\ref{eq:mfev}) evaluate to
\begin{equation}\label{eq:dnls}
\dot{z_j} = -i H^0_{jk} z_k - i \frac{\vert z_j \vert^2 z_j}{\vert {\bf z} \vert^2}.
\end{equation}
The $n \to \infty$ limit therefore produces $\dot{\bf z}$ evolution given by (\ref{eq:dnls}). This evolution conserves $\vert {\bf z} \vert$, so the $\vert {\bf z} \vert^{-2}$ factor in (\ref{eq:dnls}) is just a constant, and (\ref{eq:dnls}) can be identified as a discrete nonlinear Schr\"{o}dinger equation (DNSE). More specifically, discretizing the time-dependent Gross--Pitaevskii equation in space leads to evolution in the form of (\ref{eq:dnls}). The next step is to find suitable $\hat{X}_m$ operators. When these are chosen to make $D_\perp({\bf r})$ positive semi-definite, the stochastic quantum system of (\ref{eq:psisde}) is associated with the (\ref{eq:SDE}) SDE. In contrast to bosonic mean-field approximations, this is an exact correspondence for all $n$ and $t \geq 0$, but the DNSE is replaced by (\ref{eq:SDE}), which has the DNSE dynamics plus stochastic corrections that become small for large $n$.

The interaction Hamiltonian given by (\ref{eq:intH}) and (\ref{eq:HDNLS}) has a simple structure: there are no couplings between particle components. This motivates selecting $X_m$ with similar structure:
\begin{equation}\label{eq:DX}
X_{mjk} = \sqrt{c} \, \delta_{mj} \delta_{jk},
\end{equation}
where $c \geq 0$. Then the $\hat{X}_m$ operators are given by
\begin{equation}\label{eq:DXM}
\hat{X}_m = \sqrt{c} \, \hat{a}^\dagger_m \hat{a}_m,
\end{equation}
and there are $N$ of them. (\ref{eq:Dfull}) becomes
\begin{equation}\label{eq:GPD}
D({\bf z}) = \frac{1}{n}\begin{pmatrix}
\frac14\text{Im}\left[Z({\bf z})^2\right] + c\text{Im}\left[Z({\bf z})\right]^2 & -\frac14\text{Re}\left[Z({\bf z})^2\right]-\frac{c}{2}\text{Im}\left[Z({\bf z})^2\right] \\
-\frac14\text{Re}\left[Z({\bf z})^2\right]-\frac{c}{2}\text{Im}\left[Z({\bf z})^2\right] & -\frac14\text{Im}\left[Z({\bf z})^2\right] + c\text{Re}\left[Z({\bf z})\right]^2
\end{pmatrix},
\end{equation}
where $Z({\bf z})$ is an $N \times N$ matrix with entries
\begin{equation}
Z_{jk}({\bf z}) = \delta_{jk} z_j.
\end{equation}
Since $Z({\bf z})$ is diagonal, $D({\bf z})$ decomposes into $2 \times 2$ subspaces, each associated with the real and imaginary parts of a single ${\bf z}$ component and having the form
\begin{equation}\label{eq:Dblock}
D_j({\bf z}) = \frac{1}{n}\begin{pmatrix}
\frac14\text{Im}\left(z_j^2\right) + c\text{Im}\left(z_j\right)^2 & -\frac14\text{Re}\left(z_j^2\right)-\frac{c}{2}\text{Im}\left(z_j^2\right) \\
-\frac14\text{Re}\left(z_j^2\right)-\frac{c}{2}\text{Im}\left(z_j^2\right) & -\frac14\text{Im}\left(z^2_j\right) + c\text{Re}\left(z_j\right)^2
\end{pmatrix}.
\end{equation}
The eigenvalues of (\ref{eq:Dblock}) are
\begin{equation}
\lambda_{\pm} = \frac{\vert z_j \vert^2}{4n} \left( 2 c \pm \sqrt{1 + 4 c^2}\right).
\end{equation}
For all $c \geq 0$, $\lambda_+$ is positive and $\lambda_-$ is negative, although $\lambda_-$ does approach zero for large $c$:
\begin{equation}\label{eq:lamm}
\vert \lambda_- \vert \leq \frac{\vert z_j \vert^2}{16 n c}.
\end{equation}
The reason why $\lambda_-$ cannot be made positive is that there is an eigenvector with a negative eigenvalue that is partially along $\binom{x_j}{y_j}$, while the positive diffusion proportional to $c$ is orthogonal to $\binom{x_j}{y_j}$. This is a demonstration of the point made earlier, that $D({\bf z})$ must be replaced by $D_\perp({\bf z})$ in order to eliminate all negative diffusion. However, in this case, $D({\bf z})$ is much easier to work with than $D_\perp({\bf z})$. The projection step in (\ref{eq:Dperp}) causes the loss of the $D({\bf z})$ structure, such that $D_\perp({\bf z})$ does not simply decompose into $2 \times 2$ subspaces. Then to make $D_\perp({\bf z})$ positive semi-definite, a more complicated set of $X_m$ is required. But that can be avoided by instead using the results from Sect.~\ref{sec:apx}. (\ref{eq:lamm}) can be applied to bound the absolute sum of negative eigenvalues of the $D({\bf z})$ in (\ref{eq:GPD}), yielding
\begin{equation}
\alpha({\bf z}) \leq \sum_j \frac{\vert z_j \vert^2}{16 n c} \leq \frac{1}{16 n c}.
\end{equation}
Next, using (\ref{eq:nerel}) and (\ref{eq:frhod}), a bound of
\begin{equation}
\left\Vert \rho^+(t) - \rho(t) \right\Vert_* \leq \frac{3}{8} \frac{t}{c}
\end{equation}
is obtained for the difference between the $\rho(t)$ of the quantum evolution [given by (\ref{eq:phoev}) or (\ref{eq:psisde})] and $\rho^+(t)$, which is given by (\ref{eq:hyp}) with $f({\bf z}, t)$ being the probability distribution of the SDE of (\ref{eq:Ito}) with $D^+_\perp({\bf r})$ in place of $D({\bf r})$. Then by choosing
\begin{equation}\label{eq:c}
c = \frac{3}{8} \frac{t}{\epsilon}
\end{equation}
for some final time $t$, the quantum evolution yields a $\rho(t)$ that approximates $\rho^+(t)$ to within $\epsilon$ in trace norm. So, by (\ref{eq:outdb}), output quantities evaluated on this SDE are given to within $\mathcal{O}(\epsilon)$ by evaluating the same output on the quantum system.

\subsection{Requirements for efficiency}\label{sec:req}

Efficient simulation of the stochastic quantum systems still presents a number of challenges, which we now discuss in general, with the system from Sect.~\ref{sec:DNSE} occasionally serving as an example. The evolution in (\ref{eq:psisde}) can be approximated using finite steps $\Delta t$ in time:
\begin{equation}
\vert \psi\left(t + \Delta t)\right) \rangle = \exp\left[-i \left(\hat{H} + \sqrt{\frac{2}{\Delta t}} \sum_m Q_m \hat{X}_m\right)\Delta t\right] \vert \psi(t) \rangle,
\end{equation}
where each $Q_m$ is a random variable sampled from a distribution centered on zero with a standard deviation of one. For example, each $Q_m$ can be selected with unbiased sampling from $\pm 1$. Due to the cancelations in (\ref{eq:rhosplit}), the local errors in $\rho$ are $\mathcal{O}\left[(\Delta t)^2\right]$. Furthermore, the evolution can be split up into pieces [e.g.,  (\ref{eq:splitx})] involving subsets of the $Q_m$ random variables without affecting this error bound. The global errors are $\mathcal{O}(t \Delta t)$, in common with standard Eulerian time stepping. Then to ensure that $\rho(t)$ has error bounded by $\epsilon$, time steps with
\begin{equation}\label{eq:Dt}
\Delta t = \mathcal{O}\left(\frac{\epsilon}{t}\right)
\end{equation}
are required. This assumes that the strength of the evolution operators does not scale with $t$ or $\epsilon$, which is the normal situation. But for the particular approximation made in Sect.~\ref{sec:DNSE}, (\ref{eq:DXM}) and (\ref{eq:c}) break this assumption, and (\ref{eq:Dt}) has to be replaced with $\Delta t = \mathcal{O}(\epsilon^3/t^3)$ to ensure $\rho(t)$ error bounded by $\epsilon$.

The $\hat{X}_m$ operators have the same form as the non-interacting terms in the Hamiltonian [cf. (\ref{eq:phoev})]. Therefore, the evolution due to the $\hat{X}_m$ operators over a single time step is equivalent to some non-interacting evolution, which can be implemented by evolving each single particle state $\vert \varphi_1 \rangle$ according to
\begin{equation}\label{eq:1PE}
\frac{d}{dt} \vert \varphi_1 \rangle = -i X \vert \varphi_1 \rangle
\end{equation}
for time $\Delta t$, where $X$ is a Hermitian matrix. Specifically,
\begin{equation}\label{eq:X1}
X = \sqrt{\frac{2}{n \Delta t}} \sum_{m} Q_m X_{m}
\end{equation}
to implement the evolution by all the $\hat{X}_m$, while the evolution due to the non-interacting part of $\hat{H}$ is implemented with $X = H^0$. Either way, this is a Hamiltonian simulation problem over an $N$-dimensional Hilbert space. The $X$ matrix must have significant structure to allow efficient simulation for large $N$. For instance, $X$ can be a sparse matrix with entries that are efficiently computable based on the matrix indices \cite{Berry2016}. Hamiltonian simulation complexity also scales with a norm of the Hamiltonian. For example, \cite{Low2017} gives query complexity linear in the max norm, which translates to being linear in $\Vert X \Vert_{\text{max}}$. The $\sqrt{1/\Delta t}$ factor in (\ref{eq:X1}) generally causes some inefficiency, worsening the scaling with $t$ and $\epsilon$ on account of (\ref{eq:Dt}). This is avoidable in one special case, when the evolution due to $X$ can be efficiently fast forwarded.

For the Sect.~\ref{sec:DNSE} system, (\ref{eq:DX}) results in a diagonal $X$ [given by (\ref{eq:X1})], so an explicit solution to (\ref{eq:1PE}) is easily obtained and can be implemented directly rather than requiring a Hamiltonian simulation algorithm. Even so, efficiency is not yet ensured, because the $Q_m$ random variables still require handling. As there are $N$ of these in this case, it is not possible to generate and apply them without incurring costs linear in $N$, which we want to avoid. Therefore, we propose that the $Q_m$ instead be evaluated using a pseudorandom number generating scheme that allows efficient computation based on an index. For instance, with linear congruential generators and linear-feedback shift registers, the $m$\textsuperscript{th} number in the sequence can be found efficiently. Another possibility is to apply hash functions to $m$ to generate the pseudorandom numbers. In any case, the generated values should change for each time step and overall algorithm run. That can be achieved by combining $m$, the time step index, and the run index into a single index that is used to generate the $Q_m$ numbers.

Next, we discuss other potential sources of inefficiency that are not related to the stochastic evolution component. The initial state must be prepared. Assuming that the SDE starts in a definite state ${\bf z}_0$, the corresponding initial quantum state,
\begin{equation}\label{eq:psi0}
\vert \psi(0) \rangle = \vert {\bf z}_0 \rangle^{\otimes n},
\end{equation}
can be prepared by preparing the $\vert \psi(0) \rangle$ state $n$ times in separate registers. There is, however, no guarantee that $\vert \psi(0) \rangle$ can be prepared efficiently. Indeed, if ${\bf z}_0$ is generic, then it is not possible to avoid $\Omega(N)$ gate complexity, where $N$ is the size of ${\bf z}_0$, when preparing $\vert \psi(0) \rangle$. Still, many states can be prepared much more efficiently, including with $\text{poly}(\log N)$ complexity (e.g., \cite{Soklakov2006, Grover2002}); this is just the usual state preparation problem that is present for many quantum algorithms.

Other challenges include performing the Hamiltonian simulation efficiently and extracting the output efficiently. In both cases the associated operators need significant structure to allow a $\text{poly}(\log N)$ complexity scaling. The Hamiltonian, given in (\ref{eq:phoev}), consists of a non-interacting component and a two-particle interaction. The former is covered by the previous discussion since it corresponds to (\ref{eq:1PE}) with $X = H^0$. Next, with particles represented in separate registers, such as in (\ref{eq:psi0}), the interaction decomposes into a sum of $\mathcal{O}(n^2)$ terms over pairs of particles. The evolution due to any such term is based on $H_{jklm}$ applied to the associated particle pair $\vert \varphi_2 \rangle$, and efficiency is dependent on the structure of $H_{jklm}$. For the DNSE, $H_{jklm}$ is given by (\ref{eq:HDNLS}), and the evolution for each particle pair is just
\begin{equation}
\frac{d}{dt} \vert \varphi_2 \rangle = -i \left( \sum_j \vert j \rangle \langle j \vert \otimes \vert j \rangle \langle j \vert \right) \vert \varphi_2 \rangle,
\end{equation}
which can be implemented efficiently by applying phase factors controlled on equality between the register index bits. Next, whether an output can be extracted efficiently depends on the structure of the observable $\hat{O}$. Suppose that $\hat{O}$ is encoded as the upper-left block of a larger matrix $U$ which is unitary. Then
\begin{equation}
\text{Tr}\left(\hat{O} \vert \psi(t) \rangle \langle \psi(t) \vert\right) = \left( \vert 0 \rangle_a \vert \psi(t) \rangle \right)^\dagger U \vert 0 \rangle_a \vert \psi(t) \rangle = \left(\vert 0 \rangle_a \vert 0 \rangle_b \right)^\dagger C^\dagger(t) U C(t) \vert 0 \rangle_a \vert 0 \rangle_b,
\end{equation}
where $\vert 0 \rangle_a$ is an ancilla register, $\vert 0 \rangle_b$ is the primary register, and $C(t)$ is the quantum circuit that prepares $\vert \psi(t) \rangle$ from $\vert 0 \rangle_b$. If $\vert \psi(t) \rangle$ can be prepared efficiently and $U$ can be applied efficiently, then $C^\dagger(t) U C(t)$ can be applied efficiently. Following that circuit with measurement allows for approximating the $\hat{O}$ observable. Specifically, $\vert \text{Tr}(\hat{O} \vert \psi(t) \rangle \langle \psi(t) \vert) \vert$ can be estimated with accuracy $\epsilon$ by repeating the circuit and measurement steps $\mathcal{O}(1/\epsilon^2)$ times. Notably, the strategy of encoding a Hermitian operator as a block of a unitary is used in \cite{Qubitization} for efficient Hamiltonian simulation of a wide variety of Hamiltonians. Therefore, whether an observable can be efficiently evaluated by this method, not including the cost to prepare $\vert \psi(t) \rangle$, is roughly the same question as whether $\hat{H} = \hat{O}$ can be efficiently simulated.

We deem the overall simulation efficient if its complexity is
\begin{equation}
\mathcal{O}\left[\text{poly}\left(t, 1/\epsilon, n, \log N\right)\right].
\end{equation}
This is obtainable for the stochastic DNSE of Sect.~\ref{sec:DNSE} provided that the preparation of $\vert {\bf z}(0) \rangle$, the non-interacting evolution of (\ref{eq:1PE}) with $X = H^0$, and the unitary $U$ with the observable $\hat{O}$ encoded in its upper-left block are all implemented with $\mathcal{O}[\text{poly}(\log N)]$ complexity. More cannot be said without making specific choices for ${\bf z}(0)$, $H^0$, and $\hat{O}$. Still, this example does suggest that the simulation of some non-trivial stochastic nonlinear differential equations can be efficiently approximated on a quantum computer.

\section{Conclusion}\label{sec:conc}

We have detailed a correspondence between open bosonic quantum systems with two-particle interactions and stochastic nonlinear differential equations, with the latter being expressible as stochastic dynamical systems of $2N$ real variables, where $N$ is the number of states of each bosonic particle. This correspondence is valid when the interactions between the bosonic system and the environment are sufficiently strong, such that entanglement between the bosonic particles does not develop. As the number of bosons $n$ increases, the strength of the stochastic terms decreases, and in the $n \to \infty$ limit, a bosonic mean-field theory is recovered. At finite $n$, a stochastic mean-field theory is given, which could be applied to compute or analyze the quantum system dynamics accurately without requiring $n \gg 1$. When $N$ is not very large, the stochastic mean-field theory can be simulated efficiently on a classical computer using standard techniques such as the Euler--Maruyama method. This is particularly interesting in the $n \gtrsim N$ regime, since classical simulation of quantum systems with large $n$ is very expensive in general.

In the other direction, the correspondence allows for some stochastic nonlinear differential equations to be simulated on a quantum computer, which is potentially useful in the $n \ll N$ regime. It is hard to see how any classical algorithms for simulating a stochastic system of $2N$ real variables could avoid complexity at least linear in $N$, especially when this evolution includes nonlinear deterministic terms. Yet, provided that the conditions discussed in Sect.~\ref{sec:req} are met, there is the possibility to formulate a quantum algorithm for obtaining simulation outputs with complexity scaling as $\mathcal{O}[\text{poly}(\log N)]$ with $N$, which appears to be a significant quantum speedup relative to any classical simulation algorithm.  Note that representing bosonic particles with, e.g., the qubits of a quantum computer generally requires significant entanglement even when there is no entanglement between the original bosonic particles. We leave the task of detailing, analyzing, and testing a quantum algorithm to speed up the simulation of a stochastic nonlinear differential equation for potential future work.

\backmatter

\bmhead{Acknowledgments}
This work was supported in part by the U.S. Department of Energy under Grant No. DE-SC0020393.

\bmhead{Data Availability Statement}
Data sharing not applicable to this article as no datasets were generated or analysed during the current study.

\bibliography{bib}

\end{document}